\documentclass[aps,prl,twocolumn,amsmath,amssymb,floatfix,superscriptaddress]{revtex4}
\usepackage{tabularx}
\usepackage{bm}
\usepackage{euscript}
\usepackage{epsfig,psfrag,subfigure}
\usepackage{graphicx}
\usepackage{color}
\usepackage{amsfonts}
\usepackage{exscale}
\usepackage{amsbsy}
\usepackage{stmaryrd}
\usepackage{wasysym}

\def\avg#1{\left\langle#1\right\rangle}

\def\be{\begin{equation}}       \def\ee{\end{equation}}
\def\bea{\begin{eqnarray}}      \def\eea{\end{eqnarray}}
\def\ba{\begin{array} }
\def\ea{\end{array} }
\def\bnum{\begin{enumerate} }
\def\enum{\end{enumerate}}

\def\nn{\nonumber}
\def\pa{\partial}
\def\=>{\Rightarrow}
\def\>{\rightarrow}

\def\eye2{Fathbb{I}}

\def\Eq#1{Eq.~(\ref{#1})}
\def\Fig#1{Fig.~\ref{#1}}

\begin{document}

%\centerline{ \bf
\title{Gapless Spin Liquids: Stability and Possible Experimental Relevance}
%\title{\bf Which spin liquid is it?}
\author{Maissam Barkeshli}
\affiliation{Department of Physics, Stanford University, Stanford, California 94305, USA}
\author{Hong Yao}
\affiliation{Institute for Advanced Study, Tsinghua University, Beijing, 100084, China}
\affiliation{Department of Physics, Stanford University, Stanford, California 94305, USA}
\author{Steven A. Kivelson}
\affiliation{Department of Physics, Stanford University, Stanford, California 94305, USA}

%\date{}                                           % Activate to display a given date or no date

\date{\today }
\begin{abstract}
For certain crystalline systems, most notably the organic compound EtMe$_3$Sb[Pd(dmit)$_2$]$_2$,
experimental evidence  has accumulated of an insulating state with a high density of gapless neutral excitations that produce
Fermi-liquid-like power laws in thermodynamic quantities and thermal transport. This has been taken as evidence of a 
fractionalized spin liquid state. In this paper, we argue that
{\it if the experiments are taken at face value}, the most promising spin liquid candidates are a  $Z_4$ spin liquid with a pseudo-Fermi surface and no broken symmetries, or a $Z_2$ spin-liquid with  a
pseudo-Fermi surface and at least one of the following spontaneously broken: (a) time-reversal and inversion, (b) translation, or (c) certain point-group symmetries. We present a solvable model on the triangular lattice with an (a) type $Z_2$ spin liquid groundstate. 
%of the first of these 
%We find consistency of all these states with magneto-thermal transport measurements. 
\end{abstract}
\maketitle

The notion of a ``spin-liquid phase'' --  a quantum disordered insulating phase which is not adiabatically connected to a band
insulator -- has captured the imagination of theorists for decades \cite{Anderson1,BZA}. 
 In recent years, several developments have increased interest in this subject \cite{balents2010}, 
including a number of exact results for solvable models which prove the existence of spin liquids as theoretically stable
quantum states of matter \cite{RK, Kitaev,balents2002, YZK, YF},
numerical studies \cite{Assad, HW, YB} of more physically realistic models, increasingly sophisticated
field theoretic analyses \cite{LNW,wen04}, and, importantly, recent experimental results.
Specifically, a number of quasi-two-dimensional (2D) insulating materials have been found to exhibit highly
unusual low temperature thermodynamic and transport properties which are unlike those expected of conventional
phases \cite{YoungLee,Kanoda08,Matsuda10,balents2010}.

One of the most notable examples is the organic material EtMe$_3$Sb[Pd(dmit)$_2$]$_2$ (which we will refer to as ``dmit''), which provides a
physical realization of a frustrated spin-1/2 system on a (anisotropic) triangular lattice. Although dmit has an odd number of electrons per unit cell,
the charge response (conductivity) is insulating and NMR studies indicate that no magnetic ordering occurs down
to the lowest achievable temperatures, which are much smaller than the
scale of the characteristic exchange coupling, $J\approx 250 K$ \cite{itou2010}.
The specific heat, $C$, uniform susceptibility, $\chi$, and thermal conductivity, $\kappa_{xx}$, exhibit T dependencies consistent with Fermi-liquid-like
power laws, $C \sim k_B^2 \rho T$, $\chi \sim \rho \mu_B^2$, and $\kappa_{xx} \sim C$,
with an apparent density of states, $\rho \sim 0.1 J^{-1}$ per unit cell \cite{kato2011,Matsuda10}.
This is suggestive of the existence of a spin-liquid with a charge gap, and neutral, fermionic spinon excitations
with a non-zero density of states at zero energy and an estimated mean free path $\sim 0.5 \; \mu m$ \cite{Matsuda10}. Similar observations in a closely related compound, $\kappa$-(ET)$_2$Cu$_2$(CN)$_3$ (referred to as $\kappa$CN),\cite{shimizu2003} led to an early proposal for a candidate spin liquid, with a ``pseudo Fermi surface'' of
spinons and an emergent $U(1)$ gauge field.\cite{motrunich2005} 
Eventually it was found that  $\kappa$CN has a gap to mobile excitations,\cite{yamashita2009} 
but this does not appear to be the case in dmit.

In this paper, we \it assume \rm that dmit realizes a fractionalized spin liquid state, and
we address the problem of identifying the most promising candidates to explain the experiments.
In particular, we assume that the experimental claims summarized above can be taken at
face value, and we also consider the possibility that quenched disorder plays no fundamental role.
In agreement with previous discussions \cite{motrunich2005}, we argue that under these assumptions the 
$U(1)$ spin liquid is not a viable candidate. %We argue that the simplest spin liquid candidate that best agrees with the reported experimental

Firstly, we introduce a $Z_2$ spin liquid phase with a stable pseudo-Fermi surface {\it and} spontaneously broken time-reversal and inversion symmetry. 
Other attractive candidates with  pseudo-Fermi surfaces which are only marginally unstable are a
 $Z_4$ spin liquid with no necessary broken symmetries, or a $Z_2$ spin liquid which spontaneously breaks
 translation symmetry \cite{lee2007}, or certain point-group symmetries.
In particular, we %argue that these spin liquid
have studied the magnetic field response of these spin liquids, and conclude that these states are compatible with existing
magneto-thermal transport measurements. If weak quenched disorder does play a fundamental role, 
then other spin liquid states may also be viable candidates.\cite{Trivedi, Subir}
The broken symmetries of the $Z_2$ spin liquids with a pseudo-Fermi surface %, which are necessary for the state to be stabilized,
imply %the existence of 
at least one thermal phase transition.

\it ``Parent'' $U(1)$ spin liquid \rm-- A physically motivated starting point for the discussion of spin liquids \cite{LNW}
begins with the representation of the spin operator as $\vec{S} = \Psi^\dagger \vec{\underline{\tau}} \Psi$,
where $\Psi$ is a two-component spinor field (corresponding to the two polarizations of a spin-1/2 fermion),
and $\underline\tau_\alpha$ are the Pauli matrices. This leads to a minimal model of a sea of spinons coupled
to an emergent $U(1)$ gauge field, given by the Euclidean Lagrangian:
\bea
L = \Psi^\dagger\left[\partial_t -ia_0-\underline\epsilon(-i\vec\nabla - \vec a)\right]\Psi + \frac {|f|^2} {2g} + \ldots,
\eea
$\underline\epsilon = \epsilon_0 \underline 1+ \sum_\alpha J_\alpha \underline \tau_\alpha$ is a $2\times 2$ matrix;
time reversal symmetry implies that $\epsilon_0(\vec k) = \epsilon_0(-\vec k)$ and
$J_\alpha(\vec k) = -J_\alpha(-\vec k)$, and, in the absence of spin-orbit coupling, spin-rotational symmetry
implies $J_\alpha=0$.  Here, $\vec k$ signifies the Bloch wave-vector and, where
there are multiple spinon bands, it is implicitly assumed to include a band index. $f_{\mu\nu}$ is the field strength
corresponding to the emergent gauge field, $a_\mu$, and $\ldots$ represents four-fermion and higher order interaction terms,
subject to the constraint of spinon number conservation which is implied by the emergent gauge symmetry.  $L$ is proposed
to describe physics at energies small compared to the charge-gap, so the remaining degrees of freedom carry zero electro-magnetic charge.

The weak-coupling ($g\to 0$) fixed point is unstable.  One possible result is a strong-coupling non-Fermi-liquid
phase which does not break any symmetries, does not have any well defined quasiparticles, but which preserves the Fermi surface. % and all the same low-energy degrees of freedom.
\cite{lee1992}  However, assuming this to be a stable phase, it cannot be responsible
for the experimental claims summarized above. It would exhibit power-laws (for example, $C \sim T^{2/3}$) that are substantially different from
those of a Fermi-liquid, unless a broad intermediate finite-temperature regime is assumed that is
governed by the unstable ($g \to 0$) fixed point. Given that the instability is strongly relevant
and there are no naturally small parameters in the problem, we find such a broad intermediate regime to be unlikely.
%Even ignoring the gauge fluctuations, we can completely rule
Additional issues that cast serious doubt on the $U(1)$ spin liquid below 1 K on the basis of the thermal Hall data were presented in Refs. \cite{katsura2010,Matsuda10}.

\it Breaking $U(1)$ to $Z_2$ \rm -- The gauge symmetry in Eq. (1) can be spontaneously broken due to pairing of spinons, gapping the gauge fluctuations and leaving
a residual $Z_2$ gauge symmetry. Because  a $Z_2$ gauge theory has no finite temperature transition in two dimensions, spinon pairing defines a crossover rather than a phase transition.  The discreteness of the residual gauge symmetry implies that there are gapped, vortex-like excitations (``visons''\cite{Senthilfisher})
which carry half a quantum of gauge flux. Visons are time-reversal invariant objects - a point to which we will return below.
Now, the spinon number is only  conserved mod 2, so new terms
in the effective field theory are permitted.

At energies low compared to the vison creation energy %, and assuming %(for now)  that translation symmetry is not broken,
the resulting effective field theory is of the form: 
%\bea
%\label{Lsupercond}
$L = \Psi^\dagger\big[\partial_t+\underline\epsilon(-i\vec\nabla )+\underline \delta(-i\vec \nabla)\big]\Psi 
+\Psi^\dagger\underline\Delta(-i\vec\nabla)\Psi^\dagger
+\Psi\underline\Delta^\dag(-i\vec\nabla)\Psi+\cdots$.  
%\nonumber
%\eea
The induced changes in the dispersion, $\underline\delta(\vec k)$, can be absorbed into a redefined spinon ``band structure,''
$\tilde {\underline\epsilon}(\vec k) =\underline \epsilon(\vec k) +\underline \delta(\vec k)$.

One possibility is that pairing fully gaps the Fermi surface, leading to a class of gapped, topologically ordered,
spin liquid states, which are also ground states of many known solvable microscopic models \cite{wen04}. These states appear to be irrelevant for
describing the physics of dmit, which appears to have delocalized  gapless modes.

\it Nodal $Z_2$ spin liquids \rm -- If the resulting lines of gap nodes intersect the spinon Fermi surface, the result is a nodal spin liquid:
the gauge fluctuations are gapped, but the spinon spectrum consists of a number of gapless Dirac-like nodal points.
Dirac nodes are known\cite{Berg} to be perturbatively stable as long as time-reversal symmetry is preserved. There are various
exactly solvable models that exhibit such a stable nodal $Z_2$ spin liquid state \cite{Kitaev,YZK}

While this class of states has gapless spinon excitations, it has a vanishing density of states at zero energy, and so, it is not
a candidate to explain the experiments if the effects of disorder are negligible. It has been noted, however, that weak disorder broadens
the nodes, resulting in a constant density of states at zero energy proportional to the spinon scattering rate, $1/\tau$.  
Thus, only if the disorder plays a significant role in the thermodynamics can a nodal spin-liquid be a candidate to explain
the experiments.

\it $Z_2$ spin-liquids with a pseudo-Fermi surface
\rm 
%-- It is possible to find fine-tuned circumstances in which time-reversal symmetry is not broken, but in which the gap function, $\underline\Delta$, is such that a line of gap nodes coincides with all or a portion of the Fermi surface.  For instance, in the absence of spin-orbit coupling ($\vec J_\alpha=0$), this could happen if $\underline \Delta(\vec k) \propto \epsilon_0(\vec k)$.\cite{BZA}   The result would be a $Z_2$ spin liquid with a finite density of  states for spinons at zero energy.  This system would behave as a non-interacting Fermi gas in all ways, other than its charge response, and would thus be an excellent candidate for explaining the experimental observations in dmit. Moreover, it was found that for the $\Gamma$ matrix model on the square lattice, the ground state has just such a structure  for special values of the coupling constants.\cite{YZK}  However, such a state is unstable to gapped or nodal states. 
%
-- In the $U(1)$ spinon Fermi surface state, arbitrary pairing terms cannot appear in the theory as they are forbidden by the $U(1)$ gauge symmetry.
However once the $U(1)$ is broken to $Z_2$, the spinon number is only conserved mod 2, so any pairing terms can generically appear.
A small $s$-wave pairing term, which is allowed by symmetry, can thus open a gap everywhere, while p-wave or d-wave pairing terms
will gap the Fermi surface everywhere but at isolated nodal points. Indeed, in Ref. \cite{YZK}, the ``$\Gamma$ matrix model'' was found to have a $Z_2$ spin liquid groundstate which, under fine-tuned conditions, has  a pseudo-Fermi surface.  However, 
any change in the parameters away from this multicritical point results in a nodal spin-liquid.%\cite{YZK}

The instability inherent in a state with a spinon Fermi surface with a $Z_2$ gauge field is removed if the
system in question breaks both time-reversal and inversion symmetry.  In this case, the  degeneracy
of states at $\vec k$ and $-\vec k$ is lifted. (See Fig. 1.)  %, and in particular the Fermi surface is distorted so that it is asymmetric in $\vec k$.
If we draw the Fermi surface corresponding to $\tilde\epsilon(-\vec k)$,  (dashed line in Fig. 1b) %and then ask what happens when a small pairing term is added to the low energy effective action, we find that a gap opens only in the neighborhood of any  isolated points at which the two copies of
a small pairing term will open gaps only in the neighborhood of the points at which  the two copies of the Fermi surface happen to cross. % each other. %, as shown in Fig. 1.
The rest of the Fermi surface is perturbatively stable!

\begin{figure}[tb]
\subfigure[]{
\includegraphics[scale=0.29]{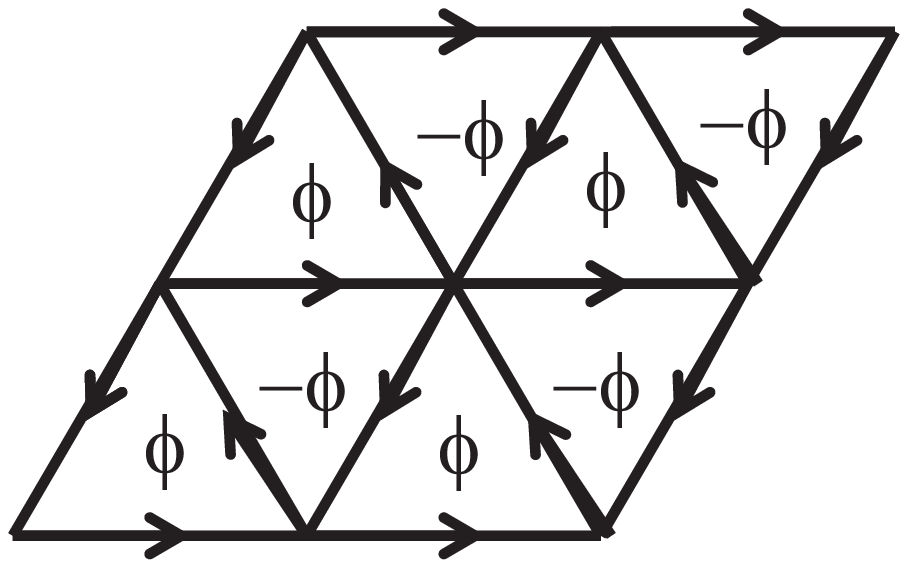}
}
\subfigure[]{
\includegraphics[scale=0.18]{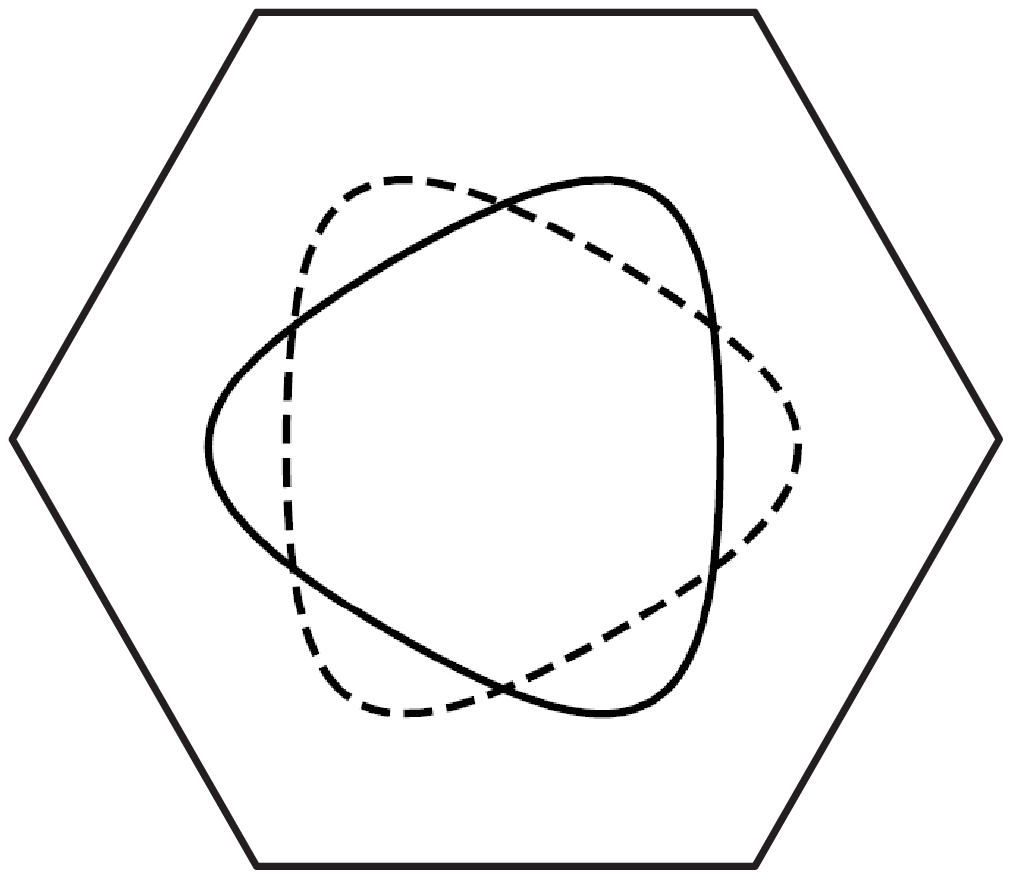}
}
\subfigure[]{
\includegraphics[scale=0.14]{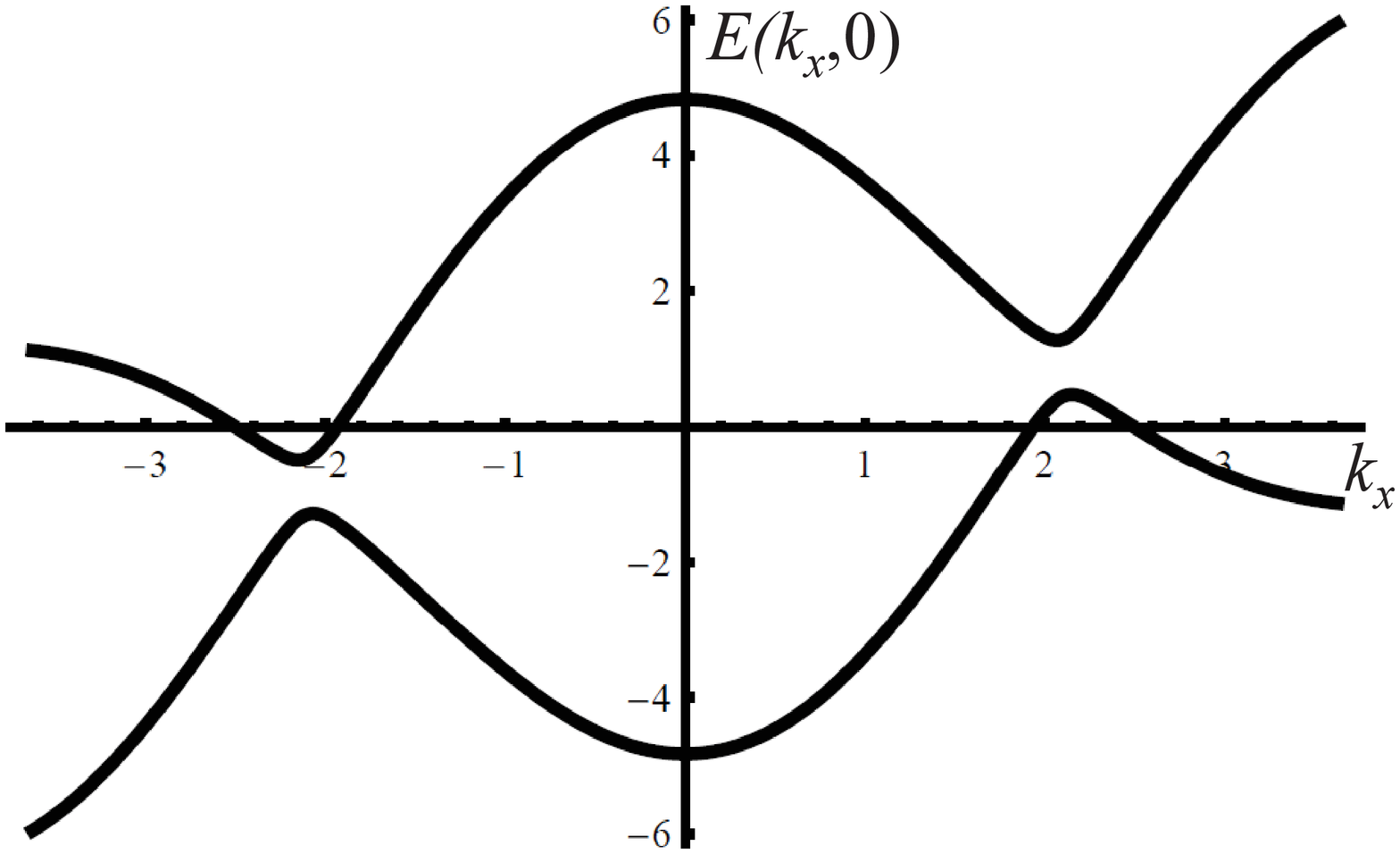}
}
\caption{{(a) Loop order on the triangular lattice where $\phi$ represents the phase accumulated in circling the plaquette. This leads to the dispersion relation $\tilde \epsilon(\vec k) = -t [\cos(k_x) + \cos(k_+)+\cos(k_-) ] - \delta [\sin(k_x) + \sin(k_+)+\sin(k_-) ] $ where $k_{\pm} = - k_x/2 \pm\sqrt{3}k_y/2$ and $\delta/t = \tan(\phi/3)$.
(b) The Fermi surface with  $\phi=\pi/2$ (solid line) and the same curve with $\vec k \to -\vec k$ (dashed line). (c) The dispersion for  $k_y=0$ in the presence of a small pairing term.}}
\label{fig:fig1}
\end{figure}

Therefore, if both time-reversal and inversion symmetry are broken, it is possible to stabilize a $Z_2$ spin liquid with
a pseudo-Fermi surface. Such a state was found in \cite{YF} in a $\Gamma$-matrix
model on the Kagome lattice with explicit time-reversal and inversion symmetry breaking.  A related result was obtained  in another context:
 in \cite{Berg} it was shown that a Fermi surface occurs in a state with coexisting d-wave superconducting and orbital loop orders.\cite{varma} %The loop order
An example of such a state on a triangular lattice is shown in Fig. 1. % leads to the dispersion %$\epsilon_0(\vec k) = -t [\cos(k_x) + \cos(-k_x/2 +\sqrt{3}k_y/2)+\cos(k_x/2 +\sqrt{3}k_y/2) ] - \delta [\sin(k_x) + \sin(-k_x/2 +\sqrt{3}k_y/2)-\sin(k_x/2 +\sqrt{3}k_y/2) ] $ where $\delta/t = \tan(\phi/3)$.
%Now, even in the presence of a non-vanishing generic pair-field, $\Delta(\vec k)$, a spinon Fermi surface is stable so long as $\Delta$ is not too large in magnitude.\cite{Berg}
A subtle point is the possibility that the spinon band structure violates time-reversal and inversion, while
physical (gauge invariant) quantities preserve these symmetries \cite{wen2002}. However such states are believed to 
be unstable to translation symmetry breaking that gaps the Fermi surface. 

To affirm the possibility of a stable $Z_2$ state with pseudo-Fermi surface, we introduce 
(in the Supplemental Material) an exactly solvable spin-7/2 version of the $\Gamma$-matrix model
on the triangular lattice.  The model itself is not inversion symmetric.  We find that for a range of parameters, the ground-state spontaneously
breaks time-reversal % and inversion symmetry,
symmetry and supports a stable pseudo Fermi surface coupled to a $Z_2$ gauge field.
%The existence of a spontaneously broken discrete symmetry in the ground-state implies the necessity of at least one  finite temperature phase transition.  {\it This establishes the existence of a stable spin-liquid phase of matter with a pseudo Fermi surface.}  This phase is a viable candidate to account for experiments, even in the clean limit.

Another possibility is a pair density wave (PDW) state for the spinons. A specific 
version of this was proposed in \cite{lee2007}. The essential feature \cite{orgad2008} of this state which prevents it from fully gapping the pseudo-Fermi surface is that the gap parameter changes sign under translation by 1/2 the PDW period, so that the spatial averaged gap vanishes.  In the special case where the period is two lattice spacings, the PDW does not actually break translation symmetry, since translation by one lattice constant is equivalent to a uniform gauge transformation.   On the triangular lattice such a state does, however, break the $C_6$ rotational symmetry.  (However,  on a square lattice, a period 2 PDW state could occur without breaking any symmetries.)  In these cases, the pseudo-Fermi surface has a marginally relevant Cooper instability.

%Such states can support a pseudo-Fermi surface %\cite{orgad2008}, albeit with a marginally relevant Cooper instability, and would break translation symmetry.

\it $Z_4$ spin liquids \rm -- We can also imagine cases where quartets condense but not pairs,
leaving  a $Z_4$ gauge theory with a richer collection of vortex-like modes.
While quartet condensation is relatively unnatural for electrons, with their strong repulsive interactions, there
is no particular reason to rule out condensation of higher multiplets in the case of spinons.  It is, for example, possible to construct
model electron problems with strong, spin-dependent attractions which exhibit a charge 4e superconducting phase.\cite{largeJ}

A charge 4e condensate can also occur if a pair-density wave state\cite{berg2007,himeda} (PDW) is quantum or thermally melted by the proliferation of double dislocations, and similar considerations apply to  FFLO states with no phase twist.\cite{radzihovsky2009}
Thermal melting of a PDW can proceed in several ways, but if it occurs by the proliferation of double dislocations, the result is a spatially uniform charge 4e superconductor.  While the theory of the quantum melting has not been worked out, presumably the proliferation of double dislocation loops would similarly lead to a charge 4e superconducting phase.  Moreover, by analogy with the similar problem of stripe melting considered in \cite{senthil2012}, we expect that
the Fermi surface of gapless quasiparticles will survive this transition intact.
  In the spin-liquid context, such a state would be a $Z_4$ spin-liquid with a pseudo Fermi surface.

A state with a $Z_4$ condensate is exotic, in the sense that there is no simple (quadratic in fermion operators) mean-field description possible.  Certainly, there are no gapless gauge fluctuations and the ``visons''
necessarily break time-reversal symmetry - a vison which carries $\phi_0/4$ flux is distinct
from its time-reversed anti-vison which carries $-\phi_0/4$.  If the
quartet binding energy is large enough, the spinon spectrum will be fully gapped.  However,
the melted PDW provides a tangible  example of a $Z_4$ liquid with a Fermi surface.
 Moreover, such a  spin liquid would be at most marginally unstable ({\it i.e.} that there is a
Cooper instability to a spinon-paired state.)  Any potentially relevant pair-field perturbation is forbidden by the residual $Z_4$ gauge
symmetry.
A $Z_4$ spin liquid is an attractive candidate for accounting for the experiments - in contrast to the $Z_2$
spin liquid, it does not necessitate time-reversal and inversion symmetry breaking to ensure the stability of the spinon-Fermi surface
over a broad intermediate energy scale. % possibly ranging to well below the lowest achievable temperatures.

\it Response to magnetic field \rm --
%Thermal Hall effect.  Kerr effect.  Coupling between magnetic fields and visons.  Absence of coupling in the quantum dimer model.
Experiments on dmit have reported no measurable thermal Hall angle up to an applied field of 12 T \cite{Matsuda10}, and an interesting
upturn in $\kappa_{xx}(T = 0)$ starting at an applied field of 2 T.

The orbital coupling to charge fluctuations leads to a linear coupling
between the magnetic field and the spin chirality.\cite{motrunich2006}
The spin chirality is proportional to the magnetic flux of the emergent gauge field.\cite{LNW}
%The gapped
However, the visons of $Z_2$ spin liquids typically %do not have any excitations that are odd
are even under time-reversal, %though this is a possibility depending on the strength of
due to tunneling between $\phi_0/2$ and $-\phi_0/2$ vortices. For strong enough tunneling, vortices are therefore not induced by a magnetic field, even if it is larger than the vison gap.  If the tunneling amplitude is sufficiently small, an  applied magnetic field could mix the symmetric and antisymmetric states, stabilizing vortices relative to anti-vortices.
When the $Z_2$ gauge field is coupled to gapless spinons, this tunneling may be further suppressed,
and could even vanish \cite{visontun}. %However i

If the effective penetration length of the spinon condensate is small (i.e. if it forms a Type I superconductor),
then we expect that a magnetic field will not induce a finite density of vortices, and therefore there
is no mechanism for modifying the thermal Hall effect. If the spinon condensate forms a Type II superconductor, then a magnetic-field
dependent thermal Hall effect is possible in principle, although at present we do not have a theoretical estimate of the
magnitude. 

Assuming the Type I scenario, the only response of the spin liquid to the magnetic field is through the Zeeman coupling. For the
pseudo-Fermi surface states discussed in this paper, this will change the density of states and will not modify
the Hall response. However, since the $Z_2$ spin liquid with a pseudo-Fermi surface must spontaneously break time-reversal and inversion
symmetry, it should have a zero-field anomalous thermal Hall response. In the clean limit, the only contribution
to the thermal Hall conductivity comes from\cite{hallrmp} the Berry curvature of the Bloch states of the spinons:
$\kappa_{xy} = \frac{\pi^2}{3} \frac{k_B^2 T}{h} \frac{1}{2\pi} \int d^2k f_{xy}(\vec{k}) n(\vec{k})$,
where $f_{xy}(\vec{k})$ is the Berry curvature, and $n(\vec{k})$ is the occupation number of the partially filled bands. Generically, we expect:
$ \frac{1}{2\pi} \int d^2k f_{xy}(\vec{k}) n(\vec{k}) \sim 1$. %, although this can in principle also be fine-tuned to be approximately zero.
 Therefore, we estimate: $\kappa_{xy}/T \sim \frac{\pi^2}{3} \frac{k_B^2}{h} \approx 10^{-12}$  W/K$^2$  .
In order to compare with the experimental results, consider $\kappa_{xy}/T$ per layer, where the layer spacing is $d \approx 1.5$ nm:
$\kappa_{xy}/Td \approx 6 \cdot 10^{-4} $ W/K$^2$m.
Using the zero temperature intercept of the longitudinal thermal conductivity,
$\lim_{T \rightarrow 0} \kappa_{xx}/Td = 0.2 $  W/K$^2$m, we predict a Hall angle
$\tan \theta_H = \kappa_{xy}/\kappa_{xx} \approx 0.003 $.
The experimental error bars on the Hall angle were reported to be $\approx 0.05$ at $0.23$ K,
$\approx 0.02$ at $0.70$ K, and $0.003$ at 1 K and 12 T \cite{Matsuda10}. Our prediction is over an order of magnitude
less than the experimental error bars on the low temperature data and therefore appears to be consistent with experiment.

The other observed feature in the magneto-thermal transport is a rapid upturn in $\kappa_{xx}(T=0)$ at 2 T. One possible explanation
is that the tunnel splitting in the vison state  is small, so that the gap to vortex formation is approximately $1 K$, and hence magnetic fields in excess of 1 T lead to vortex proliferation. %a gapped excitation that couples to the magnetic field, with a gap of approximately $1 K$
(There is also a peak
in $\kappa_{xx}(T)/T$ at $T \approx 1 K$.) However, %the existence of such %a time-reversal odd excitation will require an onset of the
the vortex state would be expected to exhibit a thermal Hall effect
due to scattering of the spinons from vortices, and  has not been observed experimentally. % excitations off of the time-reversal odd excitation, which was not observed.
One possibility is that the experimental error bars are still too large to observe this effect.
An alternative possibility is that the upturn in $\kappa_{xx}/T$ is simply due to variations of the density of states in the spinon band structure. %Fermi surface,
%which can in principle vary rather sharply with spin polarization.

Since the $Z_4$ spin liquid with a pseudo-Fermi surface need not break time-reversal symmetry, its anomalous thermal Hall response
can be exactly zero. The structure of its vortices is richer, and in particular the $Z_4$ visons are not all time-reversal symmetric
and will therefore couple linearly to a magnetic field, leading to the possibility of a non-zero thermal Hall effect. %Nevertheless, as briefly mentioned above and discussed in detail in \cite{visontun}, this will depend on whether the spinon condensate forms a type I or type II superconductor.
However, the same quantitive issues discussed in the $Z_2$ context apply here, as well.

\it Discussion \rm -- So far, we have framed the discussion in terms of the stable spin liquids derived form the ``parent'' $U(1)$ spin liquid.
More generally, all known stable spin liquids can be understood in terms of a gauge field with gauge group $G$ coupled to matter fields.
If the spinons have a Fermi surface which is at most marginally unstable, then presumably the gauge group must be discrete.
%In order to explain the experiments, all known spin liquid fixed points will require a discrete gauge field coupled to gapless fermionic excitations.

 Nodal fermions, with dispersion $\epsilon(k) \sim k^\alpha$ are %one
 another possibility, although for the familiar case of $\alpha = 1$ %is so far the only known stable example.
 the effects of disorder must be invoked to account for a non-zero density of states.
 Moreover, in this case the thermal conductivity is theoretically expected to be ``universal'' in the sense that it does not depend on $\tau$, although it does depend on the anisotropy of the nodal dispersion;  as we will show in the Supplemental material, to account for the magnitude of the observed thermal conductivity, one must assume an extremely large anisotropy corresponding to a pairing scale that is roughly $500$ times smaller than $J$. %Unless one also imagines that this state is close to a quantum critical point, there is no obvious explanation for this small scale.\cite{****}
In contrast, neither of the proposed states with pseudo-Fermi surfaces require the existence of such an unnaturally small pairing scale.
The case of $\alpha=2$ (``quadratic band touching'') does not seem to have been considered previously;  this would produce a finite density of states, although at the expense of rendering the state marginally unstable in the presence of interactions \cite{sun2009}.  %More importantly in the present context, it would be expected to have a thermal conductivity $\kappa_{xx} \sim C\  \overline{ v^2}\ \tau \sim T^2$, where the mean-squared velocity, $\overline{v^2} \sim T$.

 As we have discussed, the pseudo-Fermi surface is completely stable in the $Z_2$ case only if time-reversal and inversion are broken, and marginally unstable if only translation and/or rotational symmetry is broken.\footnote{The $C_6$ rotational symmetry of an ideal triangular lattice is weakly broken by the crystal structure of dmit;  consequently, what would have been a sharp symmetry breaking transition would be somewhat rounded.}  Otherwise, we require a larger
gauge group, such as $Z_4$ \footnote{A $Z_3$ gauge field is another simple possibility \cite{Kitaev, balents2002}, but it does not have a simple
slave-particle construction.}. An interesting possibility is given in \cite{Subir}, which consists of both a pseudo-Fermi surface of
spin-1 fermionic excitations and a node with dispersion $\epsilon(k) \sim k^3$. This state also breaks time-reversal and inversion,
although it also requires weak disorder to exhibit power-laws consistent with experiment.

One experimental signature of the chiral $Z_2$ states %, there are additional experimental consequences. The
is that the breaking of time-reversal symmetry must occur
at a finite $T$ transition (presumably in the Ising universality class) \cite{baskaran1989}, and such a transition should be observable in any
thermodynamic quantity. %, especially the specific heat.  %The transition should be rounded by a magnetic field, so looking at thedifference, $C(T,0) - C(T,H)$ might help to exhibit the critical anomaly more crisply.
Existing specific data show no sharp anomaly, but other thermodynamic quantities, such as elastic moduli, might be more sensitive.
In the time-reversal breaking phase, various anomalous response functions should be non-zero.  For example, as we have discussed, one expects a
non-zero anomalous thermal Hall effect, although it may be small.  One also expects a spontaneous Kerr effect {\it i.e.} an anomalous Hall effect, which can be measured with exquisite sensitivity.\cite{kapitulnik2006} When spin-orbit
coupling is taken into account, there should be small magnetic fields which  might be detectable in NMR or $\mu$SR, although
since these fields are proportional both to the magnitude of the time-reversal symmetry breaking order parameter and to the
spin-orbit coupling, they may be quite small.

The marginal instability of the %$Z_4$ 
time-reversal invariant spin liquids with pseudo-Fermi surfaces has possibly interesting experimental implications;  it could account for the existence of small energy scales in the problem which could depend sensitively on minor differences between materials.  For instance, it might account for the low temperature gap inferred from thermal transport in $\kappa$CN \cite{yamashita2009}.  Moreover, %this
the $Z_4$ scenario does not require any finite temperature phase transitions.

%We conclude by emphasizing that in this paper we have \it assumed \rm that dmit realizes a spin liquid state, and we have discussed
%the most promising spin liquid candidates. %There are a number of experimental results that cast some doubt on the spin liquid interpretation and suggest either q
%There may be other possible explanations including quantum criticality and/or unknown complications due to material disorder. % are the primary explanations.
%These include (1) the upturn
%in the longitudinal thermal conductivity at 2 T coupled with the zero thermal Hall angle, which, although it has a spin liquid interpretation,
%does not appear to be a very natural one, (2) the fact that closely related materials, which differ only slightly in the anisotropy of the triangular
%lattice appear to be gapped but share some similar features such as the peak at 1 K in $\kappa_{xx}/T$, suggesting a quantum phase transition
%may be responsible for the experimental phenomenology, and (3) a number of indications that at least some Fermi liquid-like thermodynamic properties
%may originate from localized degrees of freedom.
%Nevertheless, a possible gapless fractionalized spin liquid state is a strong possibility and, if established, would be a landmark discovery.
%But it sure would be cool if this were a spin liquid!

\it Acknowledgements \rm -- This work was supported in part by the Simons Foundation (MB),  Tsinghua Startup Funds and NSF grant DMR-0904264 (HY), and NSF grant DMR-0758356 (SAK).
We also thank T. Senthil and S. Sachdev for helpful comments. 

%\bibliography{spinliquid}

\begin{thebibliography}{60}
\expandafter\ifx\csname natexlab\endcsname\relax\def\natexlab#1{#1}\fi
\expandafter\ifx\csname bibnamefont\endcsname\relax
  \def\bibnamefont#1{#1}\fi
\expandafter\ifx\csname bibfnamefont\endcsname\relax
  \def\bibfnamefont#1{#1}\fi
\expandafter\ifx\csname citenamefont\endcsname\relax
  \def\citenamefont#1{#1}\fi
\expandafter\ifx\csname url\endcsname\relax
  \def\url#1{\texttt{#1}}\fi
\expandafter\ifx\csname urlprefix\endcsname\relax\def\urlprefix{URL }\fi
\providecommand{\bibinfo}[2]{#2}
\providecommand{\eprint}[2][]{\url{#2}}

\bibitem[{\citenamefont{Anderson}(1973)}]{Anderson1}
\bibinfo{author}{\bibfnamefont{P.}~\bibnamefont{Anderson}},
  \bibinfo{journal}{Mater. Res. Bull.} \textbf{\bibinfo{volume}{8}},
  \bibinfo{pages}{153} (\bibinfo{year}{1973});
\bibinfo{author}{\bibfnamefont{P.}~\bibnamefont{Anderson}},
  \bibinfo{journal}{Science} \textbf{\bibinfo{volume}{235}},
  \bibinfo{pages}{1196} (\bibinfo{year}{1987});
\bibinfo{author}{\bibfnamefont{S.~A.} \bibnamefont{Kivelson}},
  \bibinfo{author}{\bibfnamefont{D.~S.} \bibnamefont{Rokhsar}},
  \bibnamefont{and} \bibinfo{author}{\bibfnamefont{J.}~\bibnamefont{Sethna}},
  \bibinfo{journal}{Phys. Rev. B} \textbf{\bibinfo{volume}{35}},
  \bibinfo{pages}{8865} (\bibinfo{year}{1987})

\bibitem[{\citenamefont{Baskaran et~al.}(1987)\citenamefont{Baskaran, Zhou, and
  Anderson}}]{BZA}
\bibinfo{author}{\bibfnamefont{G.}~\bibnamefont{Baskaran}},
  \bibinfo{author}{\bibfnamefont{Z.}~\bibnamefont{Zhou}}, \bibnamefont{and}
  \bibinfo{author}{\bibfnamefont{P.}~\bibnamefont{Anderson}},
  \bibinfo{journal}{Solid State Commun.} \textbf{\bibinfo{volume}{63}},
  \bibinfo{pages}{973} (\bibinfo{year}{1987}).

\bibitem[{\citenamefont{Lee et~al.}(2006)\citenamefont{Lee, Nagaosa, and
  Wen}}]{LNW}
\bibinfo{author}{\bibfnamefont{P. A.}~\bibnamefont{Lee}},
  \bibinfo{author}{\bibfnamefont{N.}~\bibnamefont{Nagaosa}}, \bibnamefont{and}
  \bibinfo{author}{\bibfnamefont{X.-G.} \bibnamefont{Wen}},
  \bibinfo{journal}{Rev. Mod. Phys.} \textbf{\bibinfo{volume}{78}},
  \bibinfo{pages}{17} (\bibinfo{year}{2006}).

\bibitem[{\citenamefont{Balents}(2010)}]{balents2010}
\bibinfo{author}{\bibfnamefont{L.}~\bibnamefont{Balents}},
  \bibinfo{journal}{Nature} \textbf{\bibinfo{volume}{464}},
  \bibinfo{pages}{199} (\bibinfo{year}{2010});
\bibinfo{author}{\bibfnamefont{P.~A.} \bibnamefont{Lee}},
  \bibinfo{journal}{Science} \textbf{\bibinfo{volume}{321}},
  \bibinfo{pages}{1306} (\bibinfo{year}{2008}).

\bibitem[{\citenamefont{Rokhsar and Kivelson}(1988)}]{RK}
\bibinfo{author}{\bibfnamefont{D.~S.} \bibnamefont{Rokhsar}} \bibnamefont{and}
  \bibinfo{author}{\bibfnamefont{S.~A.} \bibnamefont{Kivelson}},
  \bibinfo{journal}{Phys. Rev. Lett.} \textbf{\bibinfo{volume}{61}},
  \bibinfo{pages}{2376} (\bibinfo{year}{1988});
\bibinfo{author}{\bibfnamefont{R.}~\bibnamefont{Moessner}} \bibnamefont{and}
  \bibinfo{author}{\bibfnamefont{S.~L.} \bibnamefont{Sondhi}},
  \bibinfo{journal}{Phys. Rev. Lett.} \textbf{\bibinfo{volume}{86}},
  \bibinfo{pages}{1881} (\bibinfo{year}{2001}).

\bibitem[{\citenamefont{Balents et~al.}(2002)\citenamefont{Balents, Fisher, and
  Girvin}}]{balents2002}
\bibinfo{author}{\bibfnamefont{L.}~\bibnamefont{Balents}},
  \bibinfo{author}{\bibfnamefont{M.~P.~A.} \bibnamefont{Fisher}},
  \bibnamefont{and} \bibinfo{author}{\bibfnamefont{S.~M.}
  \bibnamefont{Girvin}}, \bibinfo{journal}{Phys. Rev. B}
  \textbf{\bibinfo{volume}{65}}, \bibinfo{pages}{224412}
  (\bibinfo{year}{2002});
\bibinfo{author}{\bibfnamefont{T.}~\bibnamefont{Senthil}} \bibnamefont{and}
  \bibinfo{author}{\bibfnamefont{O.}~\bibnamefont{Motrunich}},
  \bibinfo{journal}{Phys. Rev. B} \textbf{\bibinfo{volume}{66}},
  \bibinfo{pages}{205104} (\bibinfo{year}{2002}); \bibinfo{author}{\bibfnamefont{O.~I.} \bibnamefont{Motrunich}},
  \bibinfo{journal}{Phys. Rev. B} \textbf{\bibinfo{volume}{67}},
  \bibinfo{pages}{115108} (\bibinfo{year}{2003}).

\bibitem[{\citenamefont{Kitaev}(2006)}]{Kitaev}
\bibinfo{author}{\bibfnamefont{A.}~\bibnamefont{Kitaev}},
  \bibinfo{journal}{Ann. Phys.} \textbf{\bibinfo{volume}{321}},
  \bibinfo{pages}{2} (\bibinfo{year}{2006});
\bibinfo{author}{\bibfnamefont{X.-G.} \bibnamefont{Wen}},
  \bibinfo{journal}{Phys. Rev. Lett.} \textbf{\bibinfo{volume}{90}},
  \bibinfo{pages}{016803} (\bibinfo{year}{2003});
\bibinfo{author}{\bibfnamefont{M.~A.} \bibnamefont{Levin}} \bibnamefont{and}
  \bibinfo{author}{\bibfnamefont{X.-G.} \bibnamefont{Wen}},
  \bibinfo{journal}{Phys. Rev. B} \textbf{\bibinfo{volume}{71}},
  \bibinfo{pages}{045110} (\bibinfo{year}{2005});
\bibinfo{author}{\bibfnamefont{D.~F.} \bibnamefont{Schroeter}},
  \bibinfo{author}{\bibfnamefont{E.}~\bibnamefont{Kapit}},
  \bibinfo{author}{\bibfnamefont{R.}~\bibnamefont{Thomale}}, \bibnamefont{and}
  \bibinfo{author}{\bibfnamefont{M.}~\bibnamefont{Greiter}},
  \bibinfo{journal}{Phys. Rev. Lett.} \textbf{\bibinfo{volume}{99}},
  \bibinfo{pages}{097202} (\bibinfo{year}{2007});
  H. Yao and S. A. Kivelson, Phys. Rev. Lett. {\bf 99}, 247203 (2007); 
  \bibinfo{author}{\bibfnamefont{C.}~\bibnamefont{Wu}},
  \bibinfo{author}{\bibfnamefont{D.}~\bibnamefont{Arovas}}, \bibnamefont{and}
  \bibinfo{author}{\bibfnamefont{H.-H.} \bibnamefont{Hung}},
  \bibinfo{journal}{Phys. Rev. B} \textbf{\bibinfo{volume}{78}},
  \bibinfo{pages}{245121} (\bibinfo{year}{2008}); %\bibitem[{\citenamefont{Yao and Lee}(2011)}]{YL}
\bibinfo{author}{\bibfnamefont{H.}~\bibnamefont{Yao}} \bibnamefont{and}
  \bibinfo{author}{\bibfnamefont{D.-H.} \bibnamefont{Lee}},
  \bibinfo{journal}{Phys. Rev. Lett.} \textbf{\bibinfo{volume}{107}},
  \bibinfo{pages}{087205} (\bibinfo{year}{2011}).

\bibitem[{\citenamefont{Yao et~al.}(2009)\citenamefont{Yao, Zhang, and
  Kivelson}}]{YZK}
\bibinfo{author}{\bibfnamefont{H.}~\bibnamefont{Yao}},
  \bibinfo{author}{\bibfnamefont{S.-C.} \bibnamefont{Zhang}}, \bibnamefont{and}
  \bibinfo{author}{\bibfnamefont{S.~A.} \bibnamefont{Kivelson}},
  \bibinfo{journal}{Phys. Rev. Lett.} \textbf{\bibinfo{volume}{102}},
  \bibinfo{pages}{217202} (\bibinfo{year}{2009}).



\bibitem[{\citenamefont{Chua et~al.}(2011)\citenamefont{Chua, Yao, and
  Fiete}}]{YF}
\bibinfo{author}{\bibfnamefont{V.}~\bibnamefont{Chua}},
  \bibinfo{author}{\bibfnamefont{H.}~\bibnamefont{Yao}}, \bibnamefont{and}
  \bibinfo{author}{\bibfnamefont{G.~A.} \bibnamefont{Fiete}},
  \bibinfo{journal}{Phys. Rev. B} \textbf{\bibinfo{volume}{83}},
  \bibinfo{pages}{180412} (\bibinfo{year}{2011}).

\bibitem[{\citenamefont{Meng et~al.}(2010)\citenamefont{Meng, Lang, Wessel,
  Assaad, and Muramatsu}}]{Assad}
\bibinfo{author}{\bibfnamefont{Z.~Y.} \bibnamefont{Meng}},
  \bibinfo{author}{\bibfnamefont{T.~C.} \bibnamefont{Lang}},
  \bibinfo{author}{\bibfnamefont{S.}~\bibnamefont{Wessel}},
  \bibinfo{author}{\bibfnamefont{F.~F.} \bibnamefont{Assaad}},
  \bibnamefont{and}
  \bibinfo{author}{\bibfnamefont{A.}~\bibnamefont{Muramatsu}},
  \bibinfo{journal}{Nature} \textbf{\bibinfo{volume}{464}},
  \bibinfo{pages}{847} (\bibinfo{year}{2010});
\bibinfo{author}{\bibfnamefont{D.}~\bibnamefont{Zheng}},
  \bibinfo{author}{\bibfnamefont{C.}~\bibnamefont{Wu}}, \bibnamefont{and}
  \bibinfo{author}{\bibfnamefont{G.-M.} \bibnamefont{Zhang}}
  (\bibinfo{year}{2010}), \eprint{arXiv:1011.5858};
\bibinfo{author}{\bibfnamefont{S.}~\bibnamefont{Sorella}},
  \bibinfo{author}{\bibfnamefont{Y.}~\bibnamefont{Otsuka}}, \bibnamefont{and}
  \bibinfo{author}{\bibfnamefont{S.}~\bibnamefont{Yunoki}}
  (\bibinfo{year}{2012}), \eprint{arXiv:1207.1783};
\bibinfo{author}{\bibfnamefont{B.~K.} \bibnamefont{Clark}},
  \bibinfo{author}{\bibfnamefont{D.~A.} \bibnamefont{Abanin}},
  \bibnamefont{and} \bibinfo{author}{\bibfnamefont{S.~L.}
  \bibnamefont{Sondhi}}, \bibinfo{journal}{Phys. Rev. Lett.}
  \textbf{\bibinfo{volume}{107}}, \bibinfo{pages}{087204}
  (\bibinfo{year}{2011}); \bibinfo{author}{\bibfnamefont{F.}~\bibnamefont{Mezzacapo}} \bibnamefont{and}
  \bibinfo{author}{\bibfnamefont{M.}~\bibnamefont{Boninsegni}},
  \bibinfo{journal}{Phys. Rev. B} \textbf{\bibinfo{volume}{85}},
  \bibinfo{pages}{060402} (\bibinfo{year}{2012}).

\bibitem[{\citenamefont{Yan et~al.}(2011)\citenamefont{Yan, Huse, and
  White}}]{HW}
\bibinfo{author}{\bibfnamefont{S.}~\bibnamefont{Yan}},
  \bibinfo{author}{\bibfnamefont{D.~A.} \bibnamefont{Huse}}, \bibnamefont{and}
  \bibinfo{author}{\bibfnamefont{S.~R.} \bibnamefont{White}},
  \bibinfo{journal}{Science} \textbf{\bibinfo{volume}{332}},
  \bibinfo{pages}{1173} (\bibinfo{year}{2011});
\bibinfo{author}{\bibfnamefont{H.-C.} \bibnamefont{Jiang}},
  \bibinfo{author}{\bibfnamefont{Z.-Y.} \bibnamefont{Weng}}, \bibnamefont{and}
  \bibinfo{author}{\bibfnamefont{D.-S.} \bibnamefont{Sheng}},
  \bibinfo{journal}{Phys. Rev. Lett.} \textbf{\bibinfo{volume}{101}},
  \bibinfo{pages}{117203} (\bibinfo{year}{2008});
\bibinfo{author}{\bibfnamefont{S.}~\bibnamefont{Depenbrock}},
  \bibinfo{author}{\bibfnamefont{I.~P.} \bibnamefont{McCulloch}},
  \bibnamefont{and}
  \bibinfo{author}{\bibfnamefont{U.}~\bibnamefont{Schollwoeck}}
  (\bibinfo{year}{2012}), \eprint{arXiv:1205.4858}.

\bibitem[{\citenamefont{Jiang et~al.}(2012)\citenamefont{Jiang, Yao, and
  Balents}}]{YB}
\bibinfo{author}{\bibfnamefont{H.-C.} \bibnamefont{Jiang}},
  \bibinfo{author}{\bibfnamefont{H.}~\bibnamefont{Yao}}, \bibnamefont{and}
  \bibinfo{author}{\bibfnamefont{L.}~\bibnamefont{Balents}},
  \bibinfo{journal}{Phys. Rev. B} \textbf{\bibinfo{volume}{86}},
  \bibinfo{pages}{024424} (\bibinfo{year}{2012});
\bibinfo{author}{\bibfnamefont{L.}~\bibnamefont{Wang}},
  \bibinfo{author}{\bibfnamefont{Z.-C.} \bibnamefont{Gu}},
  \bibinfo{author}{\bibfnamefont{F.}~\bibnamefont{Verstraete}},
  \bibnamefont{and} \bibinfo{author}{\bibfnamefont{X.-G.} \bibnamefont{Wen}}
  (\bibinfo{year}{2011}), \eprint{arXiv:1112.3331}; \bibinfo{author}{\bibfnamefont{F.}~\bibnamefont{Figueirido}},
  \bibinfo{author}{\bibfnamefont{A.}~\bibnamefont{Karlhede}},
  \bibinfo{author}{\bibfnamefont{S.}~\bibnamefont{Kivelson}},
  \bibinfo{author}{\bibfnamefont{S.}~\bibnamefont{Sondhi}},
  \bibinfo{author}{\bibfnamefont{M.}~\bibnamefont{Rocek}}, \bibnamefont{and}
  \bibinfo{author}{\bibfnamefont{D.~S.} \bibnamefont{Rokhsar}},
  \bibinfo{journal}{Phys. Rev. B} \textbf{\bibinfo{volume}{41}},
  \bibinfo{pages}{4619} (\bibinfo{year}{1990});\bibinfo{author}{\bibfnamefont{F.}~\bibnamefont{Mezzacapo}},
  \bibinfo{journal}{Phys. Rev. B} \textbf{\bibinfo{volume}{86}},
  \bibinfo{pages}{045115} (\bibinfo{year}{2012}).

\bibitem[{\citenamefont{Wen}(2004)}]{wen04}
\bibinfo{author}{\bibfnamefont{X.-G.} \bibnamefont{Wen}},
  \emph{\bibinfo{title}{Quantum Field Theory of Many-Body Systems}}
  (\bibinfo{publisher}{Oxford Univ. Press}, \bibinfo{address}{Oxford},
  \bibinfo{year}{2004}).

\bibitem[{\citenamefont{Helton {\it et al.}}(2007)}]{YoungLee}
\bibinfo{author}{\bibfnamefont{J.~S.} \bibnamefont{Helton}} \bibnamefont{}
  \bibinfo{author}{\bibnamefont{{\it et al.}}}, \bibinfo{journal}{Phys. Rev.
  Lett.} \textbf{\bibinfo{volume}{98}}, \bibinfo{pages}{107204}
  (\bibinfo{year}{2007}).

\bibitem[{\citenamefont{Yamashita et~al.}(2008)\citenamefont{Yamashita,
  Nakazawa, Oguni, Oshima, Nojiri, Shimizu, Miyagawa, and Kanoda}}]{Kanoda08}
\bibinfo{author}{\bibfnamefont{S.}~\bibnamefont{Yamashita}},
  \bibinfo{author}{\bibfnamefont{Y.}~\bibnamefont{Nakazawa}},
  \bibinfo{author}{\bibfnamefont{M.}~\bibnamefont{Oguni}},
  \bibinfo{author}{\bibfnamefont{Y.}~\bibnamefont{Oshima}},
  \bibinfo{author}{\bibfnamefont{H.}~\bibnamefont{Nojiri}},
  \bibinfo{author}{\bibfnamefont{Y.}~\bibnamefont{Shimizu}},
  \bibinfo{author}{\bibfnamefont{K.}~\bibnamefont{Miyagawa}}, \bibnamefont{and}
  \bibinfo{author}{\bibfnamefont{K.}~\bibnamefont{Kanoda}},
  \bibinfo{journal}{Nat. Phys.} \textbf{\bibinfo{volume}{4}},
  \bibinfo{pages}{459} (\bibinfo{year}{2008}).

\bibitem[{\citenamefont{Yamashita et~al.}(2010)\citenamefont{Yamashita, Nakata,
  Senshu, Nagata, Yamamoto, Kato, Shibauchi, and Matsuda}}]{Matsuda10}
\bibinfo{author}{\bibfnamefont{M.}~\bibnamefont{Yamashita}},
  \bibinfo{author}{\bibfnamefont{N.}~\bibnamefont{Nakata}},
  \bibinfo{author}{\bibfnamefont{Y.}~\bibnamefont{Senshu}},
  \bibinfo{author}{\bibfnamefont{M.}~\bibnamefont{Nagata}},
  \bibinfo{author}{\bibfnamefont{H.}~\bibnamefont{Yamamoto}},
  \bibinfo{author}{\bibfnamefont{R.}~\bibnamefont{Kato}},
  \bibinfo{author}{\bibfnamefont{T.}~\bibnamefont{Shibauchi}},
  \bibnamefont{and} \bibinfo{author}{\bibfnamefont{Y.}~\bibnamefont{Matsuda}},
  \bibinfo{journal}{Science} \textbf{\bibinfo{volume}{328}},
  \bibinfo{pages}{1246} (\bibinfo{year}{2010}).

\bibitem[{\citenamefont{Shimizu et~al.}(2003)\citenamefont{Shimizu, Miyagawa,
  Kanoda, Maesato, and Saito}}]{shimizu2003}
\bibinfo{author}{\bibfnamefont{Y.}~\bibnamefont{Shimizu}},
  \bibinfo{author}{\bibfnamefont{K.}~\bibnamefont{Miyagawa}},
  \bibinfo{author}{\bibfnamefont{K.}~\bibnamefont{Kanoda}},
  \bibinfo{author}{\bibfnamefont{M.}~\bibnamefont{Maesato}}, \bibnamefont{and}
  \bibinfo{author}{\bibfnamefont{G.}~\bibnamefont{Saito}},
  \bibinfo{journal}{Phys. Rev. Lett.} \textbf{\bibinfo{volume}{91}},
  \bibinfo{pages}{107001} (\bibinfo{year}{2003}).

\bibitem[{\citenamefont{Itou et~al.}(2010)\citenamefont{Itou, Oyamada, Maegawa,
  and Kato}}]{itou2010}
\bibinfo{author}{\bibfnamefont{T.}~\bibnamefont{Itou}},
  \bibinfo{author}{\bibfnamefont{A.}~\bibnamefont{Oyamada}},
  \bibinfo{author}{\bibfnamefont{S.}~\bibnamefont{Maegawa}}, \bibnamefont{and}
  \bibinfo{author}{\bibfnamefont{R.}~\bibnamefont{Kato}},
  \bibinfo{journal}{Nature Phys.} \textbf{\bibinfo{volume}{6}},
  \bibinfo{pages}{673} (\bibinfo{year}{2010});
\bibinfo{author}{\bibfnamefont{T.}~\bibnamefont{Itou}},
  \bibinfo{author}{\bibfnamefont{A.}~\bibnamefont{Oyamada}},
  \bibinfo{author}{\bibfnamefont{S.}~\bibnamefont{Maegawa}},
  \bibinfo{author}{\bibfnamefont{M.}~\bibnamefont{Tamura}}, \bibnamefont{and}
  \bibinfo{author}{\bibfnamefont{R.}~\bibnamefont{Kato}},
  \bibinfo{journal}{Journal of Physics: Conference Series}
  \textbf{\bibinfo{volume}{145}}, \bibinfo{pages}{012039}
  (\bibinfo{year}{2009});
\bibinfo{author}{\bibfnamefont{T.}~\bibnamefont{Itou}},
  \bibinfo{author}{\bibfnamefont{A.}~\bibnamefont{Oyamada}},
  \bibinfo{author}{\bibfnamefont{S.}~\bibnamefont{Maegawa}},
  \bibinfo{author}{\bibfnamefont{M.}~\bibnamefont{Tamura}}, \bibnamefont{and}
  \bibinfo{author}{\bibfnamefont{R.}~\bibnamefont{Kato}},
  \bibinfo{journal}{Phys. Rev. B} \textbf{\bibinfo{volume}{77}},
  \bibinfo{pages}{104413} (\bibinfo{year}{2008}).

\bibitem[{\citenamefont{Yamashita et~al.}(2011)\citenamefont{Yamashita,
  Yamamoto, Nakazawa, Tamura, and Kato}}]{kato2011}
\bibinfo{author}{\bibfnamefont{S.}~\bibnamefont{Yamashita}},
  \bibinfo{author}{\bibfnamefont{T.}~\bibnamefont{Yamamoto}},
  \bibinfo{author}{\bibfnamefont{Y.}~\bibnamefont{Nakazawa}},
  \bibinfo{author}{\bibfnamefont{M.}~\bibnamefont{Tamura}}, \bibnamefont{and}
  \bibinfo{author}{\bibfnamefont{R.}~\bibnamefont{Kato}},
  \bibinfo{journal}{Nature Comm.} \textbf{\bibinfo{volume}{2}},
  \bibinfo{pages}{275} (\bibinfo{year}{2011}).

\bibitem[{\citenamefont{Motrunich}(2005)}]{motrunich2005}
\bibinfo{author}{\bibfnamefont{O.~I.} \bibnamefont{Motrunich}},
  \bibinfo{journal}{Phys. Rev. B} \textbf{\bibinfo{volume}{72}},
  \bibinfo{pages}{045105} (\bibinfo{year}{2005});
\bibinfo{author}{\bibfnamefont{S.-S.} \bibnamefont{Lee}} \bibnamefont{and}
  \bibinfo{author}{\bibfnamefont{P.~A.} \bibnamefont{Lee}},
  \bibinfo{journal}{Phys. Rev. Lett.} \textbf{\bibinfo{volume}{95}},
  \bibinfo{pages}{036403} (\bibinfo{year}{2005}).

\bibitem[{\citenamefont{Lee et~al.}(2007)\citenamefont{Lee, Lee, and
  Senthil}}]{lee2007}
\bibinfo{author}{\bibfnamefont{S.-S.} \bibnamefont{Lee}},
  \bibinfo{author}{\bibfnamefont{P.~A.} \bibnamefont{Lee}}, \bibnamefont{and}
  \bibinfo{author}{\bibfnamefont{T.}~\bibnamefont{Senthil}},
  \bibinfo{journal}{Phys. Rev. Lett.} \textbf{\bibinfo{volume}{98}},
  \bibinfo{pages}{067006} (\bibinfo{year}{2007}).

\bibitem[{\citenamefont{Yamashita et~al.}(2009)\citenamefont{Yamashita, Nakata,
  Kasahara, Sasaki, Yoneyama, Kobayashi, Fujimoto, Shibauchi, and
  Matsuda}}]{yamashita2009}
\bibinfo{author}{\bibfnamefont{M.}~\bibnamefont{Yamashita}},
  \bibinfo{author}{\bibfnamefont{N.}~\bibnamefont{Nakata}},
  \bibinfo{author}{\bibfnamefont{Y.}~\bibnamefont{Kasahara}},
  \bibinfo{author}{\bibfnamefont{T.}~\bibnamefont{Sasaki}},
  \bibinfo{author}{\bibfnamefont{N.}~\bibnamefont{Yoneyama}},
  \bibinfo{author}{\bibfnamefont{N.}~\bibnamefont{Kobayashi}},
  \bibinfo{author}{\bibfnamefont{S.}~\bibnamefont{Fujimoto}},
  \bibinfo{author}{\bibfnamefont{T.}~\bibnamefont{Shibauchi}},
  \bibnamefont{and} \bibinfo{author}{\bibfnamefont{Y.}~\bibnamefont{Matsuda}},
  \bibinfo{journal}{Nature Phys.} \textbf{\bibinfo{volume}{5}},
  \bibinfo{pages}{44} (\bibinfo{year}{2009}).

\bibitem[{\citenamefont{Grover et~al.}(2010)\citenamefont{Grover, Trivedi,
  Senthil, and Lee}}]{Trivedi}
\bibinfo{author}{\bibfnamefont{T.}~\bibnamefont{Grover}},
  \bibinfo{author}{\bibfnamefont{N.}~\bibnamefont{Trivedi}},
  \bibinfo{author}{\bibfnamefont{T.}~\bibnamefont{Senthil}}, \bibnamefont{and}
  \bibinfo{author}{\bibfnamefont{P.~A.} \bibnamefont{Lee}},
  \bibinfo{journal}{Phys. Rev. B} \textbf{\bibinfo{volume}{81}},
  \bibinfo{pages}{245121} (\bibinfo{year}{2010}).

\bibitem[{\citenamefont{Biswas et~al.}(2011)\citenamefont{Biswas, Fu, Laumann,
  and Sachdev}}]{Subir}
\bibinfo{author}{\bibfnamefont{R.~R.} \bibnamefont{Biswas}},
  \bibinfo{author}{\bibfnamefont{L.}~\bibnamefont{Fu}},
  \bibinfo{author}{\bibfnamefont{C.~R.} \bibnamefont{Laumann}},
  \bibnamefont{and} \bibinfo{author}{\bibfnamefont{S.}~\bibnamefont{Sachdev}},
  \bibinfo{journal}{Phys. Rev. B} \textbf{\bibinfo{volume}{84}},
  \bibinfo{pages}{085141} (\bibinfo{year}{2011}).

\bibitem[{\citenamefont{Lee and Nagaosa}(1992)}]{lee1992}
\bibinfo{author}{\bibfnamefont{P.~A.} \bibnamefont{Lee}} \bibnamefont{and}
  \bibinfo{author}{\bibfnamefont{N.}~\bibnamefont{Nagaosa}},
  \bibinfo{journal}{Phys. Rev. B} \textbf{\bibinfo{volume}{46}},
  \bibinfo{pages}{5621} (\bibinfo{year}{1992});
\bibinfo{author}{\bibfnamefont{B.~I.} \bibnamefont{Halperin}},
  \bibinfo{author}{\bibfnamefont{P.~A.} \bibnamefont{Lee}}, \bibnamefont{and}
  \bibinfo{author}{\bibfnamefont{N.}~\bibnamefont{Read}},
  \bibinfo{journal}{Phys. Rev. B} \textbf{\bibinfo{volume}{47}},
  \bibinfo{pages}{7312} (\bibinfo{year}{1993});
\bibinfo{author}{\bibfnamefont{B.~L.} \bibnamefont{Altshuler}},
  \bibinfo{author}{\bibfnamefont{L.~B.} \bibnamefont{Ioffe}}, \bibnamefont{and}
  \bibinfo{author}{\bibfnamefont{A.~J.} \bibnamefont{Millis}},
  \bibinfo{journal}{Phys. Rev. B} \textbf{\bibinfo{volume}{50}},
  \bibinfo{pages}{14048} (\bibinfo{year}{1994});
\bibinfo{author}{\bibfnamefont{C.}~\bibnamefont{Nayak}} \bibnamefont{and}
  \bibinfo{author}{\bibfnamefont{F.}~\bibnamefont{Wilczek}},
  \bibinfo{journal}{Nuclear Physics B} \textbf{\bibinfo{volume}{417}},
  \bibinfo{pages}{359 } (\bibinfo{year}{1994}), ISSN \bibinfo{issn}{0550-3213};
\bibinfo{author}{\bibfnamefont{S.}~\bibnamefont{Chakravarty}},
  \bibinfo{author}{\bibfnamefont{R.~E.} \bibnamefont{Norton}},
  \bibnamefont{and} \bibinfo{author}{\bibfnamefont{O.~F.}
  \bibnamefont{Sylju\aa{}sen}}, \bibinfo{journal}{Phys. Rev. Lett.}
  \textbf{\bibinfo{volume}{74}}, \bibinfo{pages}{1423} (\bibinfo{year}{1995});
\bibinfo{author}{\bibfnamefont{D.~F.} \bibnamefont{Mross}},
  \bibinfo{author}{\bibfnamefont{J.}~\bibnamefont{McGreevy}},
  \bibinfo{author}{\bibfnamefont{H.}~\bibnamefont{Liu}}, \bibnamefont{and}
  \bibinfo{author}{\bibfnamefont{T.}~\bibnamefont{Senthil}},
  \bibinfo{journal}{Phys. Rev. B} \textbf{\bibinfo{volume}{82}},
  \bibinfo{pages}{045121} (\bibinfo{year}{2010}).

\bibitem[{\citenamefont{Katsura et~al.}(2010)\citenamefont{Katsura, Nagaosa,
  and Lee}}]{katsura2010}
\bibinfo{author}{\bibfnamefont{H.}~\bibnamefont{Katsura}},
  \bibinfo{author}{\bibfnamefont{N.}~\bibnamefont{Nagaosa}}, \bibnamefont{and}
  \bibinfo{author}{\bibfnamefont{P.~A.} \bibnamefont{Lee}},
  \bibinfo{journal}{Phys. Rev. Lett.} \textbf{\bibinfo{volume}{104}},
  \bibinfo{pages}{066403} (\bibinfo{year}{2010}).

\bibitem[{\citenamefont{Senthil and Fisher}(2000)}]{Senthilfisher}
\bibinfo{author}{\bibfnamefont{T.}~\bibnamefont{Senthil}} \bibnamefont{and}
  \bibinfo{author}{\bibfnamefont{M.~P.~A.} \bibnamefont{Fisher}},
  \bibinfo{journal}{Phys. Rev. B} \textbf{\bibinfo{volume}{62}},
  \bibinfo{pages}{7850} (\bibinfo{year}{2000}).

\bibitem[{\citenamefont{Berg et~al.}(2008)\citenamefont{Berg, Chen, and
  Kivelson}}]{Berg}
\bibinfo{author}{\bibfnamefont{E.}~\bibnamefont{Berg}},
  \bibinfo{author}{\bibfnamefont{C.-C.} \bibnamefont{Chen}}, \bibnamefont{and}
  \bibinfo{author}{\bibfnamefont{S.~A.} \bibnamefont{Kivelson}},
\bibinfo{journal}{Phys. Rev. Lett.}
  \textbf{\bibinfo{volume}{100}}, \bibinfo{pages}{027003}
  (\bibinfo{year}{2008}).

\bibitem[{\citenamefont{Varma}(1997)}]{varma}
\bibinfo{author}{\bibfnamefont{C.~M.} \bibnamefont{Varma}},
  \bibinfo{journal}{Phys. Rev. B} \textbf{\bibinfo{volume}{55}},
  \bibinfo{pages}{14554} (\bibinfo{year}{1997}).

\bibitem[{\citenamefont{Wen}(2002)}]{wen2002}
\bibinfo{author}{\bibfnamefont{X.-G.} \bibnamefont{Wen}},
  \bibinfo{journal}{Phys. Rev. B} \textbf{\bibinfo{volume}{65}},
  \bibinfo{pages}{165113} (\bibinfo{year}{2002}).

\bibitem[{\citenamefont{Baruch and Orgad}(2008)}]{orgad2008}
\bibinfo{author}{\bibfnamefont{S.}~\bibnamefont{Baruch}} \bibnamefont{and}
  \bibinfo{author}{\bibfnamefont{D.}~\bibnamefont{Orgad}},
  \bibinfo{journal}{Phys. Rev. B} \textbf{\bibinfo{volume}{77}},
  \bibinfo{pages}{174502} (\bibinfo{year}{2008});
\bibinfo{author}{\bibfnamefont{M.}~\bibnamefont{Zelli}},
  \bibinfo{author}{\bibfnamefont{C.}~\bibnamefont{Kallin}}, \bibnamefont{and}
  \bibinfo{author}{\bibfnamefont{A.~J.} \bibnamefont{Berlinsky}},
  \bibinfo{journal}{Phys. Rev. B} \textbf{\bibinfo{volume}{84}},
  \bibinfo{pages}{174525} (\bibinfo{year}{2011}).

\bibitem[{\citenamefont{Kivelson et~al.}(1990)\citenamefont{Kivelson, Emery,
  and Lin}}]{largeJ}
\bibinfo{author}{\bibfnamefont{S.~A.} \bibnamefont{Kivelson}},
  \bibinfo{author}{\bibfnamefont{V.~J.} \bibnamefont{Emery}}, \bibnamefont{and}
  \bibinfo{author}{\bibfnamefont{H.-Q.} \bibnamefont{Lin}},
  \bibinfo{journal}{Phys. Rev. B} \textbf{\bibinfo{volume}{42}},
  \bibinfo{pages}{6523} (\bibinfo{year}{1990}).

\bibitem[{\citenamefont{{\it et al}}(2007)}]{berg2007}
\bibinfo{author}{\bibfnamefont{E. Berg} \bibnamefont{{\it et al}}},
  \bibinfo{journal}{Phys. Rev. Lett.} \textbf{\bibinfo{volume}{99}},
  \bibinfo{pages}{127003} (\bibinfo{year}{2007}).

\bibitem[{\citenamefont{Himeda et~al.}(2002)\citenamefont{Himeda, Kato, and
  Ogata}}]{himeda}
\bibinfo{author}{\bibfnamefont{A.}~\bibnamefont{Himeda}},
  \bibinfo{author}{\bibfnamefont{T.}~\bibnamefont{Kato}}, \bibnamefont{and}
  \bibinfo{author}{\bibfnamefont{M.}~\bibnamefont{Ogata}},
  \bibinfo{journal}{Phys. Rev. Lett.} \textbf{\bibinfo{volume}{88}},
  \bibinfo{pages}{117001} (\bibinfo{year}{2002}).

\bibitem[{\citenamefont{Radzihovsky and Vishwanath}(2009)}]{radzihovsky2009}
\bibinfo{author}{\bibfnamefont{L.}~\bibnamefont{Radzihovsky}} \bibnamefont{and}
  \bibinfo{author}{\bibfnamefont{A.}~\bibnamefont{Vishwanath}},
  \bibinfo{journal}{Phys. Rev. Lett.} \textbf{\bibinfo{volume}{103}},
  \bibinfo{pages}{010404} (\bibinfo{year}{2009});
\bibinfo{author}{\bibfnamefont{L.}~\bibnamefont{Radzihovsky}},
  \bibinfo{journal}{Phys. Rev. A} \textbf{\bibinfo{volume}{84}},
  \bibinfo{pages}{023611} (\bibinfo{year}{2011}).

\bibitem[{\citenamefont{Mross and Senthil}(2012)}]{senthil2012}
\bibinfo{author}{\bibfnamefont{D.}~\bibnamefont{Mross}} \bibnamefont{and}
  \bibinfo{author}{\bibfnamefont{T.}~\bibnamefont{Senthil}}
  (\bibinfo{year}{2012}), \eprint{arXiv:1207.1442}.

\bibitem[{\citenamefont{Motrunich}(2006)}]{motrunich2006}
\bibinfo{author}{\bibfnamefont{O.~I.} \bibnamefont{Motrunich}},
  \bibinfo{journal}{Phys. Rev. B} \textbf{\bibinfo{volume}{73}},
  \bibinfo{pages}{155115} (\bibinfo{year}{2006}); \bibinfo{author}{\bibfnamefont{D.}~\bibnamefont{Sen}} \bibnamefont{and}
  \bibinfo{author}{\bibfnamefont{R.}~\bibnamefont{Chitra}},
  \bibinfo{journal}{Phys. Rev. B} \textbf{\bibinfo{volume}{51}},
  \bibinfo{pages}{1922} (\bibinfo{year}{1995}).

\bibitem[{\citenamefont{Barkeshli et~al.}()\citenamefont{Barkeshli, Yao, and
  Kivelson}}]{visontun}
\bibinfo{author}{\bibfnamefont{M.}~\bibnamefont{Barkeshli}},
  \bibinfo{author}{\bibfnamefont{H.}~\bibnamefont{Yao}}, \bibnamefont{and}
  \bibinfo{author}{\bibfnamefont{S.}~\bibnamefont{Kivelson}},
  \bibinfo{journal}{to appear}  (2012).

\bibitem[{\citenamefont{Nagaosa et~al.}(2010)\citenamefont{Nagaosa, Sinova,
  Onoda, MacDonald, and Ong}}]{hallrmp}
\bibinfo{author}{\bibfnamefont{N.}~\bibnamefont{Nagaosa}},
  \bibinfo{author}{\bibfnamefont{J.}~\bibnamefont{Sinova}},
  \bibinfo{author}{\bibfnamefont{S.}~\bibnamefont{Onoda}},
  \bibinfo{author}{\bibfnamefont{A.~H.} \bibnamefont{MacDonald}},
  \bibnamefont{and} \bibinfo{author}{\bibfnamefont{N.~P.} \bibnamefont{Ong}},
  \bibinfo{journal}{Rev. Mod. Phys.} \textbf{\bibinfo{volume}{82}},
  \bibinfo{pages}{1539} (\bibinfo{year}{2010}).

\bibitem[{\citenamefont{Sun et~al.}(2009)\citenamefont{Sun, Yao, Fradkin, and
  Kivelson}}]{sun2009}
\bibinfo{author}{\bibfnamefont{K.}~\bibnamefont{Sun}},
  \bibinfo{author}{\bibfnamefont{H.}~\bibnamefont{Yao}},
  \bibinfo{author}{\bibfnamefont{E.}~\bibnamefont{Fradkin}}, \bibnamefont{and}
  \bibinfo{author}{\bibfnamefont{S.~A.} \bibnamefont{Kivelson}},
  \bibinfo{journal}{Phys. Rev. Lett.} \textbf{\bibinfo{volume}{103}},
  \bibinfo{pages}{046811} (\bibinfo{year}{2009}).

\bibitem[{\citenamefont{Baskaran}(1989)}]{baskaran1989}
\bibinfo{author}{\bibfnamefont{G.}~\bibnamefont{Baskaran}},
  \bibinfo{journal}{Phys. Rev. Lett.} \textbf{\bibinfo{volume}{63}},
  \bibinfo{pages}{2524} (\bibinfo{year}{1989}).

\bibitem[{\citenamefont{Xia et~al.}(2006)\citenamefont{Xia, Beyersdorf, Fejer,
  and Kapitulnik}}]{kapitulnik2006}
\bibinfo{author}{\bibfnamefont{J.}~\bibnamefont{Xia}},
  \bibinfo{author}{\bibfnamefont{P.~T.} \bibnamefont{Beyersdorf}},
  \bibinfo{author}{\bibfnamefont{M.~M.} \bibnamefont{Fejer}}, \bibnamefont{and}
  \bibinfo{author}{\bibfnamefont{A.}~\bibnamefont{Kapitulnik}},
  \bibinfo{journal}{Applied Physics Letters} \textbf{\bibinfo{volume}{89}},
  \bibinfo{pages}{062508 } (\bibinfo{year}{2006}). %ISSN
  %\bibinfo{issn}{0003-6951}.

\end{thebibliography}

%\newpage

\appendix
\section{Supplemental Materials}
\subsection{A. Gamma matrix model with a $Z_2$ spin liquid phase which spontaneously breaks TR symmetry and has a spinon Fermi surface}
\renewcommand{\theequation}{A\arabic{equation}}%redefine the command that creates the equation no.
\setcounter{equation}{0}%reset counter

\begin{figure}[b]
\subfigure[]{
\includegraphics[scale=0.32]{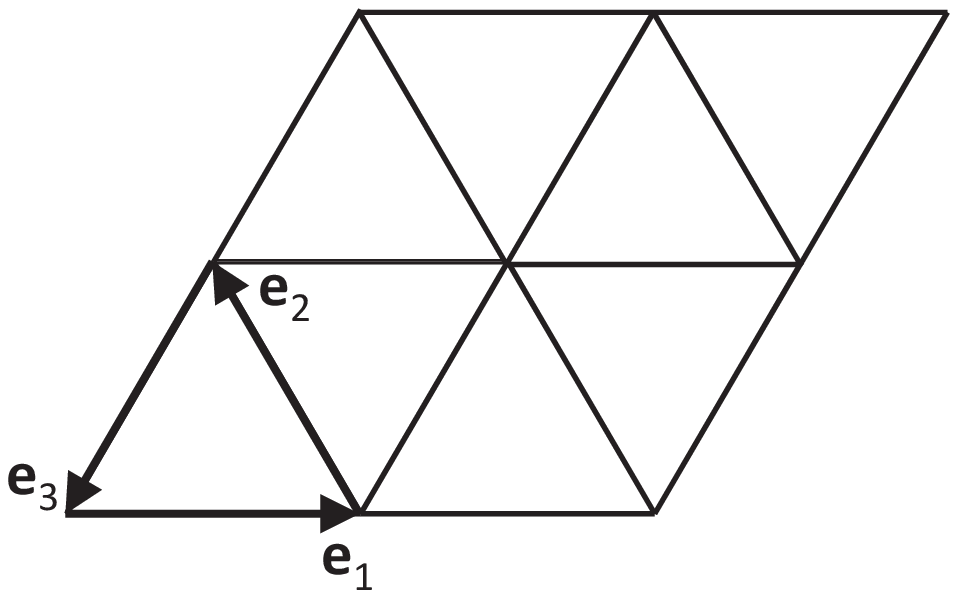}\label{fig:latt}}
\subfigure[]{
\includegraphics[scale=0.22]{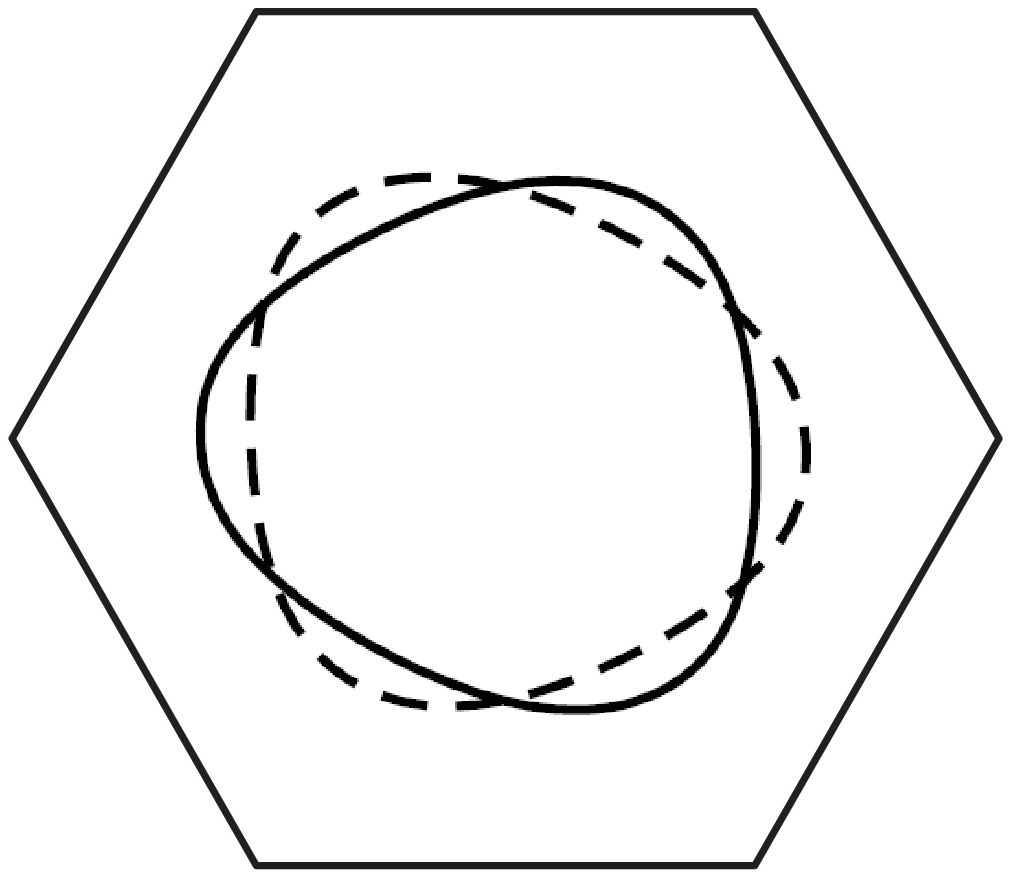}
\label{fig:fs}}
\subfigure[]{
\includegraphics[scale=0.22]{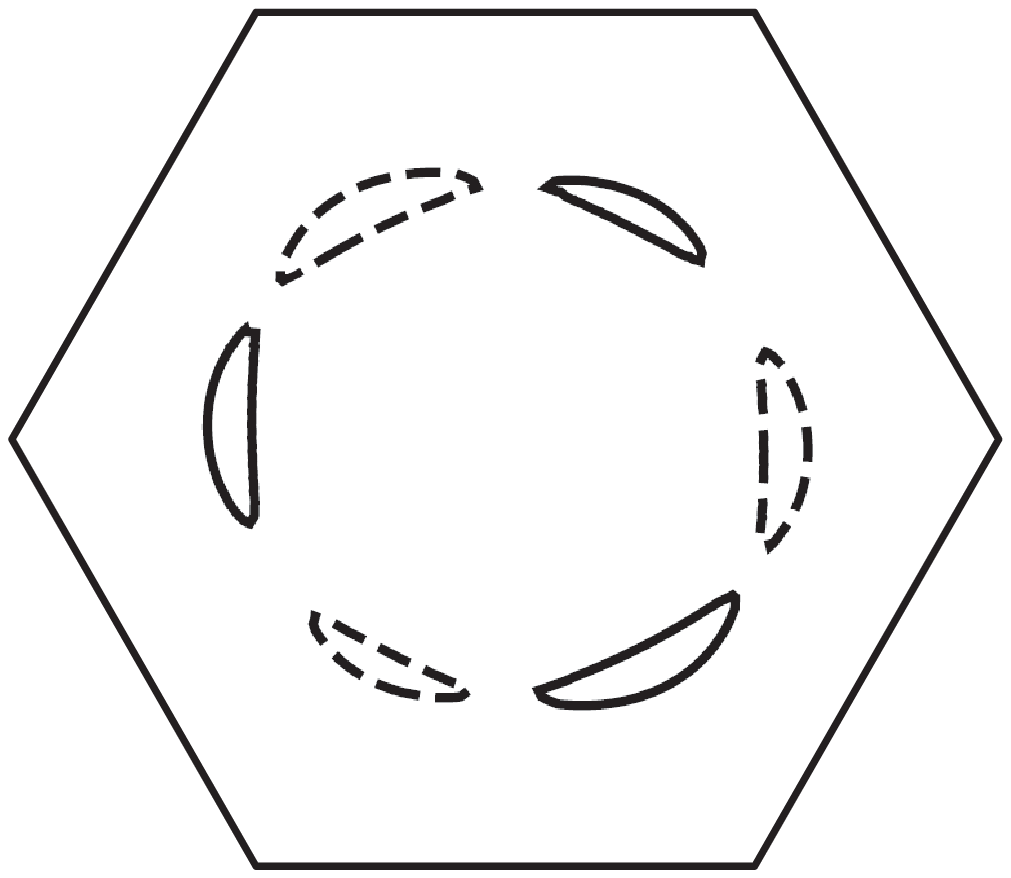}
\label{fig:fspairing}}
\caption{(a) The schematic representation of the triangular lattice. (b) The Fermi surface of $\epsilon_{\bf k}$ in \Eq{eq:bare} with  $J_\alpha=\{0.3,0.3,0.3\}$, $J'_\alpha=\{0.3,0.3,0.3\}$, $\tilde J_\alpha=\{1.0,1.0,1.0\}$, $\tilde J'_\alpha=\{-0.8,-1.0,-1.2\}$, and $J_7=0.5$. (c) The Fermi surface of $E_-(\bf k)$ in \Eq{eq:bogo}. }
\end{figure}

As a proof of principle, we consider an exactly solvable spin-7/2 model on the triangular lattice to illustrate a $Z_2$ spin liquid with a stable pseudo-Fermi surface with breaking time reversal and inversion symmetries. Following the spirit of the original $\Gamma$-matrix models\cite{YZK}, we first write down the $\Gamma$-matrices in terms of spin-7/2 operators:
\bea
\Gamma^a=f^a_{\alpha\beta\gamma} S^\alpha S^\beta S^\gamma,~ a=1,\cdots,7,
\eea
where $S^\alpha$ are the spin-7/2 operators and $f^a_{\alpha\beta\gamma}$ are a set of numbers chosen so as to satisfy the Clifford algebra of the Gamma matrices:  $\{\Gamma^a,\Gamma^b\}=2\delta^{ab}$. It is known that $f^a_{\alpha\beta\gamma}$ can always to be chosen symmetric: $f^a_{\alpha\beta\gamma}=f^a_{\beta\alpha\gamma}=f^a_{\alpha\gamma\beta}$. We denote  $\Gamma^{a,b}=[\Gamma^a,\Gamma^b]/(2i)$. Both $\Gamma^a$ and $\Gamma^{a,b}$ are odd under time reversal.

In terms of the $\Gamma$-matrices defined above, we consider the following Hamiltonian:
\bea\label{eq:ham}
H&=&\sum_{i}\sum_{\alpha=1}^3 \left[ J_\alpha \Gamma^{2\alpha-1}_i \Gamma^{2\alpha}_{i+\hat {\bf e}_\alpha} +J'_\alpha \Gamma^{2\alpha-1,7}_i \Gamma^{2\alpha,7}_{i+\hat{\bf e}_\alpha}\right.\nn\\
&+&\left.\tilde J_\alpha \Gamma^{2\alpha-1}_i \Gamma^{2\alpha,7}_{i+\hat {\bf e}_\alpha}+\tilde J'_\alpha \Gamma^{2\alpha-1,7}_i \Gamma^{2\alpha}_{i+\hat {\bf e}_\alpha}+ J_7\Gamma^7_i  \right]\nn\\ &&-K \sum_{\avg{pp'}} W_p W_{p'},
\eea
where $J_\alpha$, $J'_\alpha$, $\tilde J_\alpha$, and $\tilde J'_\alpha$ are coupling parameters, $W_p$ are the plaquette operators defined for each triangular plaquette, $\avg{pp'}$ denotes two edge-sharing triangular plaquettes. The plaquette operator $W_p$ is defined as
\bea
W_p=\Gamma^{16}_i\Gamma^{23}_{i+
\mathbf{e}_1}\Gamma^{45}_{i+\mathbf{e}_1+\mathbf{e}_2}
\eea
for up-triangles and
\bea
W_p=\Gamma^{16}_i\Gamma^{23}_{i-
\mathbf{e}_1}\Gamma^{45}_{i-\mathbf{e}_1-\mathbf{e}_2}
\eea
for down-triangular plaquettes. Note that the Hamiltonian is time reversal symmetric but explicitly  breaks inversion symmetry of the triangular lattice.

We represent the Gamma matrix operators in terms of bilinear Majorana fermions:
\bea
\Gamma^a=i\eta\gamma^a,~ \Gamma^{a,7}=i\xi\gamma^a, ~a=1,\cdots,6,
\eea
where $\eta,\xi,\gamma^a$ are Majorana fermions. This Majorana fermion representation enlarges the Hilbert space of the spins and generate a $Z_2$ gauge redundancy on each site\cite{Kitaev}. With this fermion representation,  the Hamiltonian in \Eq{eq:ham} is given by
\bea\label{eq:freeham}
H&\!=\!&\!\sum_i\!\sum_{\alpha=1}^3\!\Big[u_{i,i+{\bf e}_\alpha} \big(J_\alpha i\eta_i\eta_{i+{\bf e}_\alpha}\!+\!J'_{\alpha} i\xi_i\xi_{i+{\bf e}_\alpha}\!+\! \tilde J_\alpha i\eta_i\xi_{i+{\bf e}_\alpha}\nn\\
&+&\!\! \tilde J'_\alpha i\xi_i\eta_{i+{\bf e}_\alpha}\big) +J_7 i\eta_i\xi_i\Big]
-K \sum_{\avg{pp'}} \exp[i(\phi_p+\phi_{p'})]
\eea
where $u_{i,i+{\bf e}_\alpha}=i\gamma^{2\alpha-1}_i\gamma^{2\alpha}_{i+{\bf e}_\alpha}$ and $\exp(i\phi_p)=\prod_{\avg{jk}\in p} i u_{jk}$ (here we implicitly assumed that the order of $\avg{jk}$ in the product is taken counterclockwise). It is straightforward to show that the operators $u_{i,i+{\bf e}_\alpha}$ are conserved in the enlarged Hilbert space, which makes the model exactly solvable since the Hamiltonian in \Eq{eq:freeham} consists of quadratic fermions coupled with this static background $Z_2$ gauge field $u_{i,i+{\bf e}_\alpha}=\pm 1$. The flux on each triangular plaquette $\phi_p$ is $\pm \pi/2$.

For $K>0$ and $K\gg |J_\alpha|$, $|J'_\alpha|$, $|\tilde J_\alpha|$, $|\tilde J'_\alpha|$, and $|J_7|$, the sum of neighboring $\phi_p$ and $\phi_{p'}$ must be zero in the ground state, which implies that the flux of up-triangle plaquettes must be opposite to the down-triangles. This flux pattern spontaneously breaks time reversal symmetry, as desired. So the Fermi surface, if exists, will be stable against weak perturbations, as shown below. In the following, we shall focus on the flux sector with $\pi/2$ on up-triangles and $-\pi/2$ on down-triangles.

We combine two Majorana fermions $\eta$ and $\xi$ on each site to form one complex fermions $f$ with $\eta=f+f^\dag$ and $
\xi=i(f^\dag-f)$. Then the Hamiltonian \Eq{eq:freeham} is given by
\bea\label{eq:bdg}
H&=&\sum_i \bigg\{ \sum_{\alpha=1}^3\Big[\big[(\tilde J_\alpha-\tilde J'_\alpha)+i(J_\alpha+J'_\alpha)\big]f^\dag_if_{i+ {\bf e}_\alpha}\nn\\ \!\!\!&+&\big[(\tilde J_\alpha+\tilde J'_\alpha)+i(J_\alpha-J'_\alpha)\big]f^\dag_i f^\dag_{i+{\bf e}_\alpha}+h.c.\Big] \nn\\
 &-& 2J_7 f^\dag_i f_i \bigg\}.
\eea

The above Hamiltonian has an emergent $U(1)$ symmetry (global Fermion number conservation) when $J_\alpha=J'_\alpha$ and $\tilde J_\alpha=-\tilde J'_\alpha$, for which the pairing terms are identically zero. It is straightforward to obtain the single-particle dispersion of the hopping part:
\bea\label{eq:bare}
\epsilon_{\bf k} = 4\sum_{\alpha=1}^3 \left[J_\alpha\sin({\bf k}\cdot {\bf e}_\alpha)+\tilde J_\alpha\cos({\bf k}\cdot {\bf e}_\alpha) \right] -2J_7.~~
\eea
For $J_\alpha=\{0.3,0.3,0.3\}$, $J'_\alpha=\{0.3,0.3,0.3\}$, $\tilde J_\alpha=\{1.0,1.0,1.0\}$, $\tilde J'_\alpha=\{-0.8,-1.0,-1.2\}$, and $J_7=0.5$, the Fermi surface of $\epsilon_{\vec k}=0$ is plotted in \Fig{fig:fs}. It is clear that $\epsilon_{-{\bf k}} \neq \epsilon_{\bf k}$ for generic momentum ${\bf k}$ which is a consequence of broken time reversal and inversion symmetries. The feature makes the Fermi surface perturbatively stable in the sense that adding weak pairing term cannot fully gap the Fermi surface.

By Fourier transform, the full quadratic Hamiltonian is given by
\bea
H&=&\sum_{\bf k}\Psi^\dag_{\bf k} h_{\bf k} \Psi_{\bf k},
\eea
where $\Psi^\dag_{\bf k}=(f^\dag_{\bf k}, f_{-\bf k})$ and
\bea
h_{\bf k}=\frac12\left(\ba{cc}
\epsilon_{\bf k} & \Delta_{\bf k} \\
\Delta_{\bf k}^\ast & -\epsilon_{-\bf k}
\ea
\right).
\eea
%From \Eq{eq:bdg}, we obtain $\Delta_{\bf k}=\sum_\alpha (J_\alpha-J'_\alpha)\sin({\bf k}\cdot{\bf e}_\alpha)$.
Here $\Delta_{\bf k}=\sum_{\alpha=1}^3 2[(J_\alpha-J'_\alpha)+i(\tilde J_\alpha+\tilde J'_\alpha)]\sin({\bf k}\cdot{\bf e}_\alpha)$.
The quasi-particle spectrum is given by
\bea\label{eq:bogo}
E_{\pm}({\bf k}) = \frac{\epsilon_{\bf k}-\epsilon_{-\bf k}}2\pm \sqrt{(\frac{\epsilon_{\bf k}+\epsilon_{-\bf k}}2)^2+|\Delta_{\bf k}|^2}. 
\eea
The condition for $E_{\pm}({\bf k})=0$ is given by
\bea\label{eq:cond}
\epsilon_{\bf k}\epsilon_{-\bf k}=-|\Delta_{\bf k}|^2.
\eea
When the system breaks both time reversal and inversion symmetries, $\epsilon_{-\bf k}\neq \epsilon_{\bf k}$ in general. Then, \Eq{eq:cond} is equivalent to only one real constraint. Since there are two momenta $k_x$ and $k_y$, it is then clear that a line of zero-energy momentum points is generally allowed and stable. The Fermi surface of $E_-(\bf  k)=0$ is shown in \Fig{fig:fspairing}. (If the system were time reversal invariant which requires $\epsilon_{\bf k}=\epsilon_{-\bf k}$ and $\Delta_{\bf k}$ is real; it is then clear that \Eq{eq:cond} would imply two constraints: $\epsilon_{\bf k}=0$ and $\Delta_{\bf k}=0$, which generally allows at most nodal points, if not fully gapped.)

%We now plot the Fermi surface for a generic case with both hopping and parings in the Hamiltonian. The Fermi surface is plotted in \Fig{fig:fs2} for the case $J_\alpha=\{1.0,1.0,1.0\}$, $J'_\alpha=\{0.8,0.8,0.8\}$, $\tilde J_\alpha=\{1.0,1.0,1.0\}$, $\tilde J'_\alpha=\{0.8,0.5,1.2\}$, and $J_7=0.8$.

A $Z_2$ spin liquid with a stable pseudo-Fermi surface and broken time reversal symmetry is expected to exhibit finite thermal Hall conductivity which is linear in temperature. The thermal Hall conductivity is given by\cite{hallrmp}
\bea
\kappa_{xy}=\frac{\pi^2}{3}\frac{k_B^2 T}{h}\left[\frac{1}{2\pi}\int d^2 k f_{xy}({\bf k}) n_{\bf k}\right],
\eea
where $f_{xy}({\bf k})$ is the Berry curvature of the valence band in the momentum space and $n_{\bf k}$ is one (zero) for occupied (empty) states. We denote $h_{\bf k}=\sum_{\mu=0}^3 d_\mu({\bf k}) \sigma^\mu$, where $\sigma^0$ and $\sigma^{1,2,3}$ are identity and Pauli matrices, respectively. Then, the Berry curvature is given by
\bea
f_{xy}({\bf k})=\frac{1}{4\pi} \hat d({\bf k})\cdot \pa_{k_x} \hat d({\bf k})\times \pa_{k_y} \hat d({\bf k})
\eea
where $\hat d({\bf k}) =\vec d({\bf k})/|\vec d({\bf k})|$ and $\vec d({\bf k})=(d_1({\bf k}),d_2({\bf k}),d_3({\bf k}))$. It is clear that the Berry curvature cannot be zero everywhere when the phase of $\Delta_{\bf k}$ is not uniform in the Brillouin zone. This is fulfilled when $(J_\alpha-J'_\alpha)/(\tilde J_\alpha+\tilde J'_\alpha)$ are not all equal to one another. $J_\alpha=\{0.3,0.3,0.3\}$, $J'_\alpha=\{0.3,0.3,0.3\}$, $\tilde J_\alpha=\{1.0,1.0,1.0\}$, $\tilde J'_\alpha=\{-0.8,-1.0,-1.2\}$, and $J_7=0.5$, we obtain the thermal Hall conductivity $k_{xy}/T\approx 0.2 (\frac{\pi^2}{3}\frac{k_B^2}{h})$.

%For the specific Hamiltonian \Eq{eq:bdg}, it turns out the Berry curvature is identically zero in the entire Brillouin zone because the pairing is real which implies zero thermal Hall conductivity. But, zero Berry curvature in the entire Brillouin zone is special to the model considered in \Eq{eq:bdg} and is not generic.

%Generally, the pairing amplitude $\Delta_{\bf k}$ is complex since the system's time reversal symmetry is already spontaneously broken. To have complex pairing $\Delta_{\bf k}$, we consider the following.  In general, the Hamiltonian is given by
%\bea
%H&=&\sum_i \left[ \sum_{\alpha}\left[(J_\alpha+J'_\alpha) (if^\dag_if_{i+ {\bf e}_\alpha}+h.c.)\right.\right.\nn\\ &&\left.\left.+(J_\alpha-J'_\alpha)(if^\dag_i f^\dag_{i+{\bf e}_\alpha}+h.c.)\right] -2J_7 f^\dag_i f_i \right],\\
%&=&\sum_{\bf k}\Psi^\dag_{\bf k} h_{\bf k} \Psi_{\bf k},
%\eea
%where $\Psi^\dag_{\bf k}=(f^\dag_{\bf k}, f_{-\bf k})$ and
%\bea
%h_{\bf k}=\left(\ba{cc}
%\epsilon_{\bf k} & \Delta_{\bf k} \\
%\Delta_{\bf k}^\ast & -\epsilon_{-\bf k}
%\ea
%\right).
%\eea
%Here $\Delta_{\bf k}=\sum_\alpha (J_\alpha-J'_\alpha)\sin({\bf k}\cdot{\bf e}_\alpha)$. From my numerical calculation, it turns out the Berry curvature is identically zero even with finite pairing term. This might be related to the special case of the Kitaev model, which the hopping and pairing have the same structure.

\subsection{B. Estimate of pairing scale for nodal spin liquids}
\renewcommand{\theequation}{B\arabic{equation}}%redefine the command that creates the equation no.
\setcounter{equation}{0}%reset counter

%In this section, we use the experimental results to provide some estimates of the mean-free path of the spinons, their average velocity,
%and use these quantities to determine the conditions under which a $Z_2$ spin liquid with a nodal Dirac spinon dispersion is a viable candidate.
%\begin{comment}
%First, we expect that the mean-free velocity of the spinons is set by the exchange constant $J$, as there is no other energy scale in the
%problem (the various ring-exchange coefficients are roughly within the same order of magnitude):
%\begin{align}
%v \sim J a/\hbar \approx 3.3 \times 10^4 m/s,
%\end{align}
%where $a \approx 1 nm$ is the lattice spacing. We expect that the thermal conductivity of the spinons is approximately
%\begin{align}
%\kappa \approx \frac{1}{3} C v l,
%\end{align}
%where $C$ is the specific heat, and $l$ is the mean-free path. Setting $C = \gamma T$, this implies
%\begin{align}
%\lim_{T \rightarrow 0} \kappa/T = 1/3 \gamma v l.
%\end{align}
%From the data, $\lim_{T \rightarrow 0} \kappa/T = 0.2 W K^{-2} m^{-1}$. Thus, we estimate a mean-free path:
%\begin{align}
%l \approx 2.2 \mu m.
%\end{align}
%This implies a scattering rate
%\begin{align}
%\hbar/k_B \tau = \hbar v/l \approx 0.1 K.
%\end{align}
%\end{comment}

For a Dirac node, the universal thermal conductivity is expected to be
\begin{align}
\kappa/T = \frac{k_B^2}{3\hbar} (\frac{v_F}{v_\Delta} + \frac{v_\Delta}{v_F}),
\end{align}
where $v_F$ and $v_\Delta$ are the velocities across and along the Fermi surface, respectively, at the nodes. Since
\begin{align}
\frac{k_B^2}{3\hbar d} = 0.4 \times 10^{-3} W K^{-2} m^{-1},
\end{align}
and the measured value is $\lim_{T \rightarrow 0} \kappa/T = 0.2 W K^{-2} m^{-1}$, we see that the velocity anistropy must be
\begin{align}
v_F/v_\Delta \sim 500,
\end{align}
which is extremely large.

For a pairing scaling $\Delta$, we expect the velocity anistropy to be
\begin{align}
v_F/v_\Delta \approx J/\Delta \approx 500,
\end{align}
which requires $\Delta \approx 0.5 K$. Such an extremely small pairing scale relative to the exchange coupling $J$
requires some explanation. It has been proposed that the pairing scale is roughly $\Delta \sim 1 K$, which corresponds to
a drop in the NMR relaxation rate. The small pairing scale has been suggested to be a conseqence of proximity to a nearby quantum
critical point separating the nodal $Z_2$ spin liquid and the $U(1)$ spin liquid. Above this scale, the 
system may be well-described by a $U(1)$ spin liquid with spinon Fermi surface. A possible 
advantage of the pseudo Fermi surface proposals is that a small pairing scale is not required to explain 
the large thermal conductivity. Furthermore, if the ground state is indeed a nodal
spin liquid, another natural explanation for the small pairing scale is a marginal instability of the $Z_4$ spin liquid
with pseudo Fermi surface. 

\end{document}